\documentclass[10pt]{article}
\usepackage{amsfonts,epsfig,graphicx,subfigure,url,wrapfig,color}

\title{PSPACE-Completeness of \\Sliding-Block Puzzles and Other Problems
through the\\ Nondeterministic Constraint Logic Model of Computation}
 \author{%
      Robert A. Hearn%
         \thanks{MIT Computer Science and Artificial Intelligence Laboratory,
         	32 Vassar Street, Cambridge, MA 02139, U.S.A.,
                 \protect\url{{rah,edemaine}@csail.mit.edu}}
 \and Erik D. Demaine%
 	\footnotemark[1]
 }
\date{}

\newtheorem{thm}{Theorem}
\newtheorem{lem}[thm]{Lemma}
\newtheorem{cor}[thm]{Corollary}
\newcommand{\pf}{\noindent{\em Proof.}\hspace{1em}}

\makeatletter
\def\GrabProofArgument[#1]{ (#1): \egroup\ignorespaces}

\def\GrabProofArgument[#1]{ #1: \egroup\ignorespaces}
\def\pf{\noindent\textbf\bgroup Proof%
           \@ifnextchar[{\GrabProofArgument}{: \egroup\ignorespaces}}

\makeatother

\advance \topmargin by -\headheight
\advance \topmargin by -\headsep
\evensidemargin \oddsidemargin
\let\latexcite=\cite
\def\cite{\nolinebreak\latexcite}
\let\latexref=\ref
\def\ref{\nolinebreak\latexref}

{\makeatletter \gdef\fps@figure{!htbp}}

\newenvironment{proofsketch}{\begin{proof}[sketch]}{\end{proof}}
\let\realendproofsketch=\endproofsketch
\def\endproofsketch{\hspace*{\fill}$\Box$\realendproofsketch}

\def\fakesmallcaps{\footnotesize}

\begin{document}

\maketitle

\begin{abstract}
We present a nondeterministic model of computation based on reversing edge
directions in weighted directed graphs with minimum in-flow constraints on
vertices.  Deciding whether this simple graph model can be manipulated in order
to reverse the direction of a particular edge is shown to be PSPACE-complete by
a reduction from Quantified Boolean Formulas.  We prove this result in a
variety of special cases including planar graphs and highly restricted vertex
configurations, some of which correspond to a kind of passive constraint logic.
Our framework is inspired by (and indeed a generalization of)
the ``Generalized Rush Hour Logic'' developed by Flake and Baum
\cite{Flake-Baum-2002}.

We illustrate the importance of our model of computation by giving simple
reductions to show that several motion-planning problems are PSPACE-hard.
Our main
result along these lines is that classic unrestricted sliding-block puzzles are
PSPACE-hard, even if the pieces are restricted to be all dominoes ($1
\times 2$ blocks) and the goal is simply to move a particular piece.
No prior complexity results were known about these puzzles.
This result can be seen as a strengthening of the existing result that the
restricted Rush Hour$^{\rm TM}$
puzzles are PSPACE-complete \cite{Flake-Baum-2002},
of which we also give a simpler proof. We also greatly strengthen the conditions for
the PSPACE-hardness of the Warehouseman's Problem \cite{Hopcroft-Schwartz-Sharir-1984},
a classic motion-planning problem.
Finally, we strengthen the existing result that the pushing-blocks puzzle
Sokoban is PSPACE-complete \cite{Culberson-1998}, by showing that it is
PSPACE-complete even if no barriers are allowed.
\end{abstract}

\section{Introduction}

\begin{wrapfigure}[13]{r}{1.8 in}
  \centering
  \includegraphics[scale=0.49]{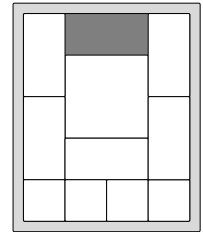}
  \caption{\label{sliding block puzzles}
    The Donkey Puzzle: move the large square to the bottom center.}
\end{wrapfigure}

\paragraph*{Motivating Application: Sliding Blocks.}

Motion planning of rigid objects is concerned with whether a
collection of objects can be moved (translated and
rotated),
without intersection among the objects,
to reach a goal configuration with certain properties.
Typically, one object
is distinguished, the remaining
objects serving as obstacles, and the goal is for
that object to reach a particular position.
This general problem arises in a variety of applied contexts such as robotics
and graphics.
In addition, this problem arises in the recreational context of
\emph{sliding-block puzzles} \cite{Hordern-1986},
where the pieces are typically integral
rectangles, L shapes, etc., and the goal is simply to move a particular piece
to a specified target position.  See Figure~\ref{sliding block puzzles} for
an example.

The \emph{Warehouseman's Problem} \cite{Hopcroft-Schwartz-Sharir-1984} is a
particular formulation of this problem in which the objects are rectangles of
arbitrary side lengths, packed inside a rectangular box.
In 1984, Hopcroft, Schwartz, and Sharir \cite{Hopcroft-Schwartz-Sharir-1984}
proved that deciding whether the rectangular objects can be moved so that
each object is at its specified final position is PSPACE-hard.
Their construction critically requires that some rectangular objects have
dimensions that are proportional to the box dimensions.

Although not mentioned in \cite{Hopcroft-Schwartz-Sharir-1984},
the Warehouseman's Problem captures a particular form of sliding-block puzzles
in which all pieces are rectangles.  However, two differences between the two
problems are that sliding-block puzzles typically require only a particular piece to
reach a position, instead of the entire configuration, and that sliding-block
puzzles involve blocks of only constant size.
More generally, it is natural to ask for the complexity of the problem
as determined by the goal specification and the set of allowed block types.

In this paper, we prove that the Warehouseman's Problem and sliding-block
puzzles are PSPACE-hard even for $1 \times 2$ rectangles (dominoes) packed
in a rectangle.  In contrast, there is a simple polynomial-time algorithm for
$1 \times 1$ rectangles packed in a rectangle.  Thus our results are
in some sense tight.

\paragraph*{Hardness Framework.}

To prove that sliding blocks and other problems are PSPACE-hard,
this paper builds a general framework for proving PSPACE-hardness
which simply requires the construction of a couple of gadgets
that can be connected together in a planar graph.
Our framework is inspired by the one developed by Flake and Baum
\cite{Flake-Baum-2002}, but is simpler and more powerful.
We prove that several different models of increasing simplicity
are equivalent, permitting simple constructions of PSPACE-hardness.
In particular, we derive simple constructions for sliding blocks,
Rush Hour \cite{Flake-Baum-2002}, and a restricted form of
Sokoban \cite{Culberson-1998}.

\paragraph*{Nondeterministic Constraint Logic Model of Computation.}

Our framework can also be viewed as a model of computation in its own right.
We show that a Nondeterministic Constraint Logic (NCL) machine has the same
computational power as a space-bounded Turing machine. Yet, it has a more
concise formal description, and has a natural interpretation as a kind
of logic network. Thus, it is reasonable to view NCL as a simple  computational
model that corresponds to the class PSPACE, just as, for example, deterministic
finite automata correspond to regular languages.

\paragraph*{Roadmap.}
Section \ref{our model} describes our model of computation in more detail.
Section \ref{Reductions} proves increasingly simple formulations of NCL to be PSPACE-complete.
Section \ref{Applications} proves various puzzles and motion-planning problems to be
PSPACE-hard using the restricted forms of our model of computation.
Section \ref{Tokens} presents an alternative formulation of the NCL problem.

\section{Nondeterministic Constraint Logic}
\label{our model}

In this section we formally define the nondeterministic constraint logic (NCL)
model of computation, and give a family of related decision problems whose complexity we are interested in.
We then describe in detail a special case of NCL graphs, called \emph{A{\fakesmallcaps ND}/O{\fakesmallcaps R} constraint graphs}.
In the following section we will show the decision problems to be PSPACE-complete,
even for the restricted \textsc{And}/\textsc{Or} constraint graphs.

\subsection{Graph Formulation}
An NCL ``machine'' is specified by a \emph{constraint graph}: an undirected
graph together with an assignment of nonnegative integers (\emph{weights})
to edges and integers (\emph{minimum in-flow constraints}) to vertices.
A configuration of this machine is an orientation (direction) of the edges
such that the sum of incoming edge weights at each vertex is at least the
minimum in-flow constraint of that vertex.
A move from one configuration to another configuration is simply the reversal
of a single edge such that the minimum in-flow constraints remain satisfied.
The standard decision question from a particular NCL machine and configuration
is whether a specified edge can be eventually reversed by a sequence of moves.
We can view such a sequence as a nondeterministic computation.

Two related decision problems are also of interest:

\begin{enumerate}
\item Given two configurations $A$ and $B$ of an NCL machine, is
there a sequence of moves from $A$ to $B$?

\item Given two edges \textsf{$E_A$} and $E_B$ of an NCL machine, and orientations for each, are there configurations
$A$ and $B$ such that $E_A$ has its desired orientation in $A$, $E_B$ has its desired orientation in $B$, and
there is a sequence of moves from $A$ to $B$?
\end{enumerate}

We refer to these three decision problems as \emph{configuration-to-edge}, \emph{configuration-to-configuration}, and
\emph{edge-to-edge}. We will show that all of these problems are PSPACE-complete. (The fourth possibility, \emph{edge-to-configuration}, 
is the same as \emph{configuration-to-edge}, by reversibility.)

\subsection{A{\fakesmallcaps ND}/O{\fakesmallcaps R} Constraint Graphs}
\label{and-or}

Certain vertex configurations in NCL graphs are of particular interest. A vertex with minimum in-flow constraint $2$ and incident edge weights of $1$, $1$, and $2$ behaves as a logical \textsc{And}, in the following sense: the weight-$2$ edge may be directed outward if and only if both weight-$1$ edges are directed inward. (Otherwise, the in-flow constraint of $2$ would not be met.) We will call such a vertex an \emph{A{\fakesmallcaps ND} vertex}.

Similarly, a vertex with minimum in-flow constraint $2$ and incident edge weights of $2$, $2$, and $2$ behaves as a logical \textsc{Or}: a given edge may be directed outward if and only if at least one of the other two edges is directed inward. We will call such a vertex an \emph{O{\fakesmallcaps R} vertex}.

\begin{figure}
  \def\gatescale{1.1}
\centering
\subfigure[\textsc{And} vertex. Edge \textsf{C} may be directed outward if and only if edges \textsf{A} and \textsf{B} are both directed inward.]
{\begin{minipage}{0.45\linewidth}
    \hfil
	\includegraphics[scale=\gatescale]{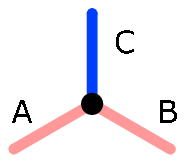}
	\label{and-vertex}
    \hfil
 \end{minipage}
}\hfil
\subfigure[\textsc{Or} vertex. Edge \textsf{C} may be directed outward if and only if either edge \textsf{A} or edge \textsf{B} is directed inward.]
{\begin{minipage}{0.45\linewidth}
    \hfil
	\includegraphics[scale=\gatescale]{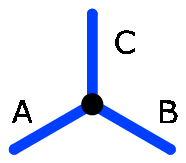}
	\label{or-vertex}
    \hfil
 \end{minipage}
}
\caption{\label{and-or-graphs} \textsc{And} and \textsc{Or} vertices. Red (light gray) edges have weight 1, blue (dark gray) edges have weight 2, and all vertices have a minimum in-flow constraint of 2.}
\end{figure}

In our reductions we will be concerned primarily with graphs containing only \textsc{And} and \textsc{Or} vertices; we call such graphs \emph{A{\fakesmallcaps ND}/O{\fakesmallcaps R} constraint graphs}.
\footnote{
We mention without proof that every NCL graph is logspace-reducible to an equivalent \textsc{And}/\textsc{Or} constraint graph (equivalent with respect to the decision problems). The reduction is rather elaborate, and the result is not needed for any of our main results, so we omit it.
} When drawing graphs, we will follow the convention that all vertices have in-flow constraint $2$, red (light gray) edges have weight $1$, and blue (dark gray) edges have weight $2$. 
Figure~\ref{and-or-graphs} shows \textsc{And} and \textsc{Or} vertices.

\paragraph*{Circuit Interpretation.}
With these \textsc{And} and \textsc{Or} vertex interpretations, it is natural to view an \textsc{And}/\textsc{Or} graph as a kind of digital logic network, or circuit. One can imagine signals flowing through the graph, as outputs activate when their input conditions are satisfied. This is the picture that motivates our description of NCL as a model of computation, rather than simply as a set of decision problems. Indeed, it is natural to expect (even before our PSPACE-completeness results) that a finite assemblage of such logic gadgets could be used to build a polynomial-space bounded computer. 

However, several differences between \textsc{And}/\textsc{Or} constraint graphs and ordinary digital logic circuits are noteworthy. First, NCL machines are inherently nondeterministic; digital logic circuits are deterministic. Second, with the above  \textsc{And} and \textsc{Or} vertex interpretations, there is nothing to prohibit ``wiring'' a vertex's ``output'' (e.g. the weight-2 edge of an \textsc{And} vertex) to another ``output'', or an ``input'' to an ``input''; in digital logic circuitry, such connections would be illegal, and meaningless. Finally, although we have \textsc{And}- and \textsc{Or}-like devices, there is nothing like an inverter (or \textsc{Not} gate)  in NCL; inverters are essential in ordinary digital logic.

This last point deserves some elaboration. The logic that is manifested in NCL graphs is a passive constraint logic; nothing forces an \textsc{And} vertex, say, to direct its weight-$2$ edge outward when its other two edges are directed inward. A signal is thus \emph{permitted}, but not required, to flow. For there to be an inverter vertex in NCL would require that an edge be permitted to be directed outward if and only if another edge is not permitted to be directed inward. But there is no way for the vertex to know, so to speak, whether an edge \emph{can} be directed inward; the constraints are in terms of what \emph{is} directed inward.

Flake and Baum require the use of inverters in a similar computational context \cite{Flake-Baum-2002}. They define gadgets (``both'' and ``either'') that are essentially the same as our \textsc{And} and \textsc{Or} vertices, but rather than use them as primitive logical elements, they use their gadgets to construct a kind of dual-rail logic. With this dual-rail logic, they can represent inverters. We do not need inverters for our reductions, so we may omit this step.

\paragraph*{Directionality; Splitting.}
As implied above, although it is natural to think of \textsc{And} and \textsc{Or} vertices as having inputs and outputs, there is nothing enforcing this interpretation. A sequence of edge reversals could first direct both red edges into an \textsc{And} vertex, and then direct its blue edge outward; in this case, we will sometimes say that its \emph{inputs} have \emph{activated}, enabling its \emph{output} to activate. But the reverse sequence could equally well occur. In this case we could view the \textsc{And} vertex as a \emph{split}: directing the blue edge inward allows both red edges to be directed outward, effectively splitting the signal.

In the case of \textsc{Or} vertices, again, we can speak of an active input enabling an output to activate. However, here the choice of input and output is entirely arbitrary, because \textsc{Or} vertices are symmetric.

\paragraph*{Red-Blue Conversion.}

\begin{figure}
\centering
\includegraphics[scale=0.9]{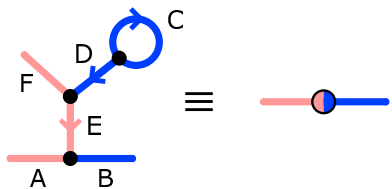}
\caption{\label{red-to-blue}
Red-to-blue conversion subgraph, with shorthand notation.%
}
\end{figure}

Viewing \textsc{And}/\textsc{Or} graphs as circuits, we might want to connect
the output of an \textsc{Or}, say, to an input of an \textsc{And}. We can't do
this directly by joining the loose ends of the two edges, because one edge is
blue and the other is red. But we can insert a
subgraph that has the desired effect, allowing the \textsc{And} input edge to
activate (point inward) just when the \textsc{Or} output edge is activated
(pointing outward). More generally, the subgraph on the left side of
Figure~\ref{red-to-blue} effectively copies a signal between a red edge and
a blue edge.
\footnote{
Our notion of graph allows loop edges and multiple edges between
   a single pair of vertices (sometimes called a \emph{multigraph} or
   \emph{pseudograph}). In Section \ref{polyhedron} we show how to reduce these graphs to simple graphs (without loops
   or multiple edges); however, this step is not strictly necessary for our applications.
 }
Edge \textsf{C} permanently satisfies the incident \textsc{Or} vertex's
constraint, allowing \textsf{D} to point away from it.
This lets \textsf{E} point down, providing an
extra in-flow of $1$ to the vertex between \textsf{A} and \textsf{B}. Either
\textsf{A} or \textsf{B} can now satisfy this vertex's constraint by pointing
inward; the other edge is free to point away.
In other words,
a signal can propagate out from
\textsf{A} precisely if it was passed in via \textsf{B}, and vice versa.

This subgraph has an extra red edge (\textsf{F}), whose direction is not constrained; we must somehow deal with its loose end.  A little reflection shows that such extra edges must always occur in pairs: red edges only exist on red-red-blue vertices, therefore the sequence \textsf{F},  \textsf{E}, \textsf{A}, must continue on in an unbranching chain, ending at another unattached red edge. We can then identify \textsf{F} with that edge, forming a cycle.

We use the shorthand notation on the right side of Figure~\ref{red-to-blue} to denote this subgraph; this will simplify the figures.

\section{PSPACE-completeness}
\label{Reductions}
In this section, we show that all three NCL decision problems are PSPACE-complete, and extend these results to apply to simplified forms of NCL.

\subsection{Nondeterministic Constraint Logic}
\label{ncl-reduction}
We show that configuration-to-edge is PSPACE-hard by giving a reduction from Quantified Boolean
Formulas (QBF),
which is known to be PSPACE-complete \cite{Garey-Johnson-1979}, even when the
formula is in conjunctive normal form (CNF).  A simple
argument then shows that configuration-to-edge is in PSPACE, and therefore PSPACE-complete.
The PSPACE-completeness of the other two decision problems also follows simply.

\begin{figure}
\centering
\includegraphics[width=.7\linewidth]{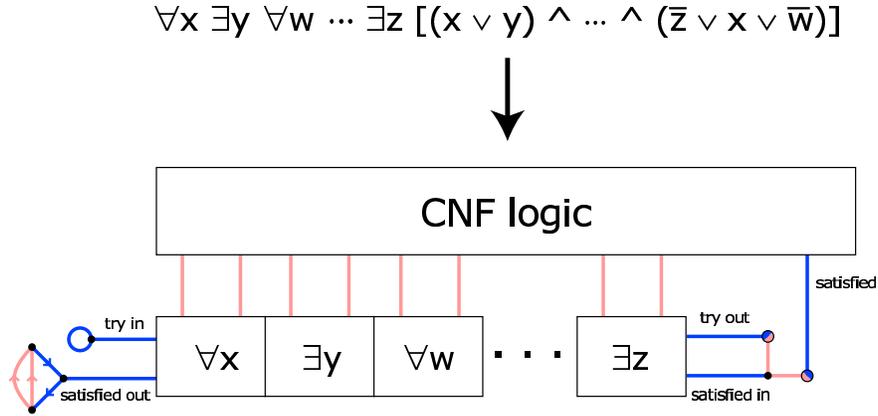}
\caption{\label{qbf}
  Schematic of the reduction from Quantified Boolean Formulas to NCL.}
\end{figure}

\paragraph*{Reduction.}
First we give an overview of the reduction and the gadgets we need; then we
analyze the gadgets' properties.

The reduction is illustrated schematically in Figure~\ref{qbf}.  We translate a given
quantified Boolean formula $\phi$ into an \textsc{And}/\textsc{Or} constraint graph, so that a particular edge in the
graph may be reversed if and only if $\phi$ is true.

One way to determine the truth of a quantified Boolean formula is as follows:  Consider the initial
quantifier in the formula.  Assign its variable first to false and then to true, and for each assignment,
recursively ask whether the remaining formula is true under that assignment.  For an existential
quantifier, return true if either assignment succeeds; for a universal quantifier, return true only
if both assignments succeed.  For the base case, all variables are assigned, and we only need to test
whether the CNF formula is true under the current assignment.

\begin{figure}
\centering
\subfigure[Quantifier gadget connections]
{%
	\includegraphics[scale= .65]{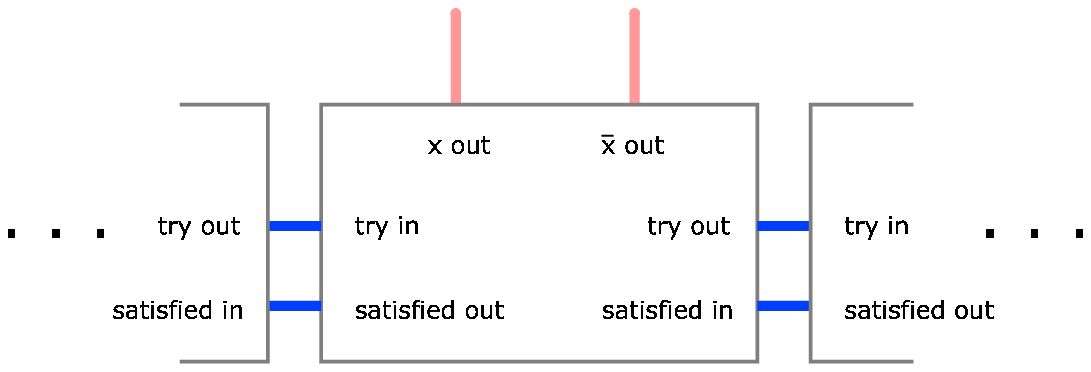}%
	\label{connections}%
}\hfill
\subfigure[Part of a CNF formula graph]
{%
	\includegraphics[scale=.5]{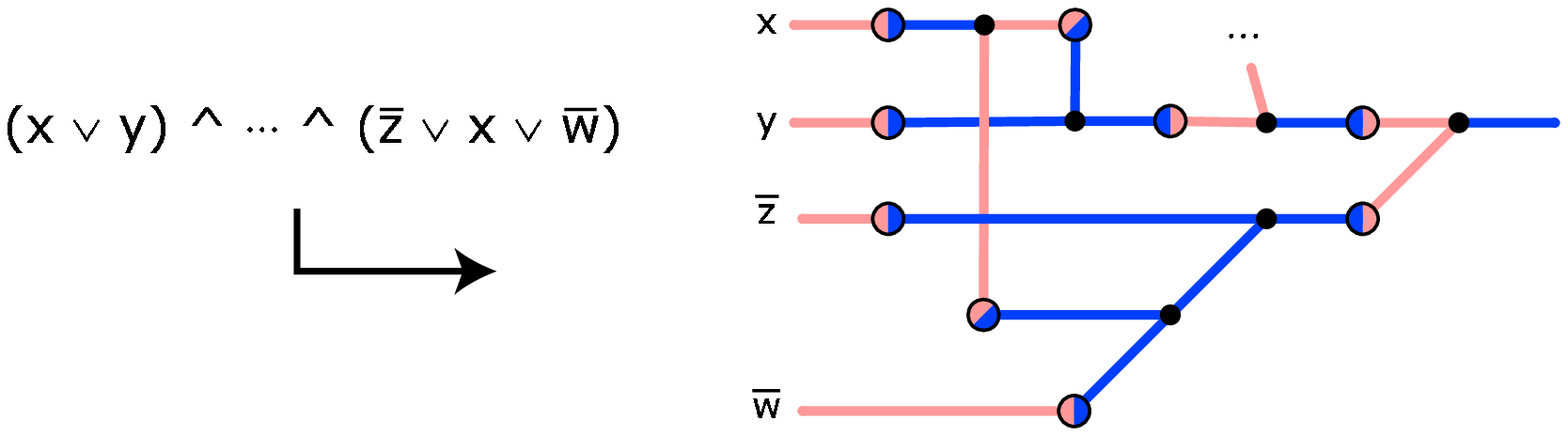}%
	\label{cnf-graph}%
}%
\caption{\label{wiring}
  QBF wiring.}
\end{figure}

This is essentially the strategy our reduction shall employ.  We define 
\emph{quantifier gadgets}, which are connected
together into a string, one per
quantifier in the formula, as in Figure~\ref{connections}. Each quantifier gadget outputs a pair of edges
corresponding to a variable assignment. These
edges feed into the \emph{CNF network}, which corresponds to the unquantified formula.
The output from the CNF network connects to
the rightmost quantifier gadget; the output of our overall graph is the \textsf{satisfied out} edge
from the leftmost quantifier gadget. (We use the attached subgraph for the other decisions problems.)

\paragraph*{Quantifier Gadgets.}

When a quantifier gadget is \emph{activated}, all quantifier gadgets to its left
have fixed particular variable assignments, and only this quantifier gadget and
those to the right are free to change their variable assignments.
The activated quantifier gadget can declare itself \emph{satisfied} if and only if
the Boolean formula read from here to the right is true given the variable
assignments on the left.

A quantifier gadget is activated by directing its \textsf{try in} edge inward.  Its \textsf{try out} edge
is enabled to be directed outward only if \textsf{try in} is directed inward, and its variable state is locked.  
A quantifier gadget may nondeterministically ``choose'' a variable assignment, and
recursively ``try'' the rest of the
formula under that assignment and those that are locked by quantifiers to its left.
The variable assignment is represented by two output edges ($x$ and $\overline{x}$), only one of which may be directed outward.
For \textsf{satisfied out} to be directed outward, indicating that the formula from this quantifier on is
currently satisfied, we require (at least) that \textsf{satisfied in} be directed inward.

We construct both existential and universal quantifier gadgets, described below, satisfying the above requirements.

\begin{lem}
\label{trysatlemma}
A quantifier gadget's \textsf{satisfied in} edge may not be directed inward unless its \textsf{try out} edge is directed outward.
\end{lem}

\begin{pf}
By induction. The condition is explicitly satisfied in the construction for the rightmost quantifier
gadget, and each quantifier gadget requires \textsf{try in} to be directed inward before \textsf{try out}
is directed outward, and requires \textsf{satisfied in} to be directed inward before \textsf{satisfied out} is directed outward.
\end{pf}

\paragraph*{CNF Formula.}

In order to evaluate the formula for a particular variable assignment, we construct an
\textsc{And}/\textsc{Or} subgraph corresponding to the unquantified part of the formula,
fed inputs from the variable gadgets, and feeding into the
\textsf{satisfied in} edge of the rightmost quantifier gadget, as in Figure~\ref{qbf}. 
The \textsf{satisfied in} edge of the rightmost quantifier gadget
is further protected by an \textsc{And} vertex, so it may be directed inward only if
\textsf{try out} is directed outward and the formula is currently satisfied.

Because the formula is in conjunctive normal form, and
we have edges representing both literal forms of each variable (true and false),
we don't need an inverter for this construction. We use the signal-splitting and red-blue
conversion techniques described in Section~\ref{and-or} to construct the graph.
Part of such a graph is shown in Figure~\ref{cnf-graph}.

\begin{lem}
\label{cnflemma}
The \textsf{satisfied out} edge of a CNF subgraph may be directed outward if and only if its corresponding formula is satisfied
by the variable assignments on its input edge orientations.
\end{lem}

\begin{pf}
Definition of \textsc{And} and \textsc{Or} vertices, and the CNF construction described.
\end{pf}

\paragraph*{Latch Gadget.}

\begin{figure}
  \def\latchscale{.75}
\centering
\subfigure[Locked, \textsf{A} active]
{\begin{minipage}{0.23\linewidth}
    \hfil
	\includegraphics[scale=\latchscale]{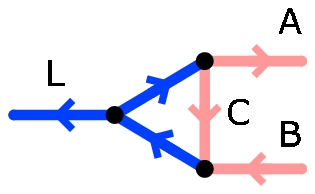}%
	\label{latch1}%
    \hfil
 \end{minipage}
}\hfil
\subfigure[Unlocked, \textsf{A} active]
{\begin{minipage}{0.23\linewidth}
    \hfil
	\includegraphics[scale=\latchscale]{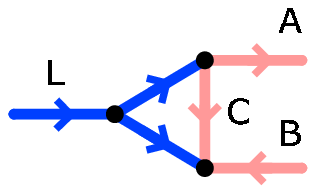}%
	\label{latch2}%
    \hfil
 \end{minipage}
}\hfil
\subfigure[Unlocked, \textsf{B} active]
{\begin{minipage}{0.23\linewidth}
    \hfil
	\includegraphics[scale=\latchscale]{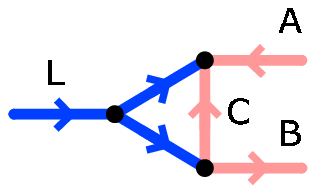}%
	\label{latch3}%
    \hfil
 \end{minipage}
}\hfil
\subfigure[Locked, \textsf{B} active]
{\begin{minipage}{0.23\linewidth}
    \hfil
	\includegraphics[scale=\latchscale]{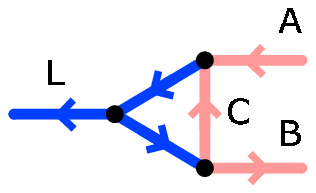}%
	\label{latch4}%
    \hfil
 \end{minipage}
}\hfil
\caption{\label{latch}
  Latch gadget, transitioning from state \textsf{A} to state \textsf{B}.}
\end{figure}

Internally, the quantifier gadgets use \emph{latch gadgets}, shown in Figure~\ref{latch}. This subgraph effectively stores a bit of information, whose state can be locked or unlocked. With edge \textsf{L} directed left, one of the other two \textsc{Or} edges must be directed inward, preventing its output red edge from pointing out. The orientation of edge \textsf{C} is fixed in this state. When \textsf{L} is directed inward, the other \textsc{Or} edges may be directed outward, and the red edges are free to reverse. Then when the latch is locked again, by directing \textsf{L} left, the state has been switched.

\begin{figure}
\centering
\subfigure[Existential quantifier]
{%
	\includegraphics[scale= .66]{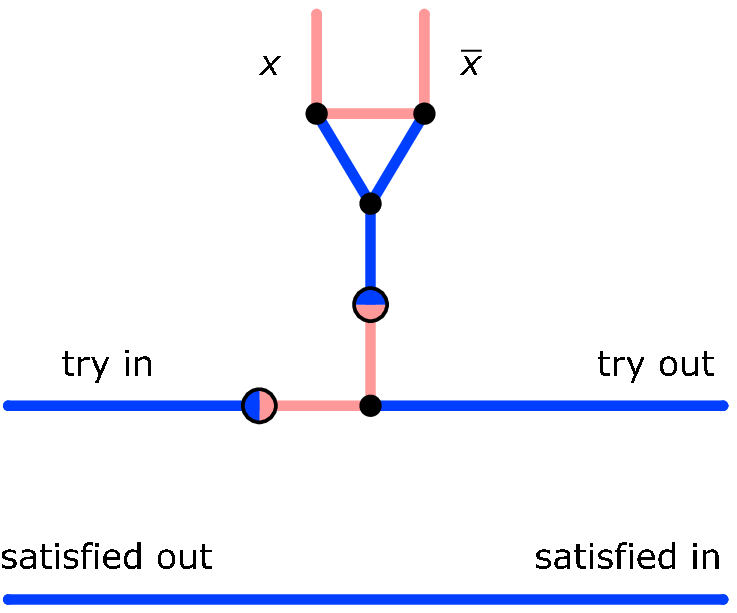}%
	\label{existential}%
}\hspace{.5 in}
\subfigure[Universal quantifier]
{%
	\includegraphics[scale=.66]{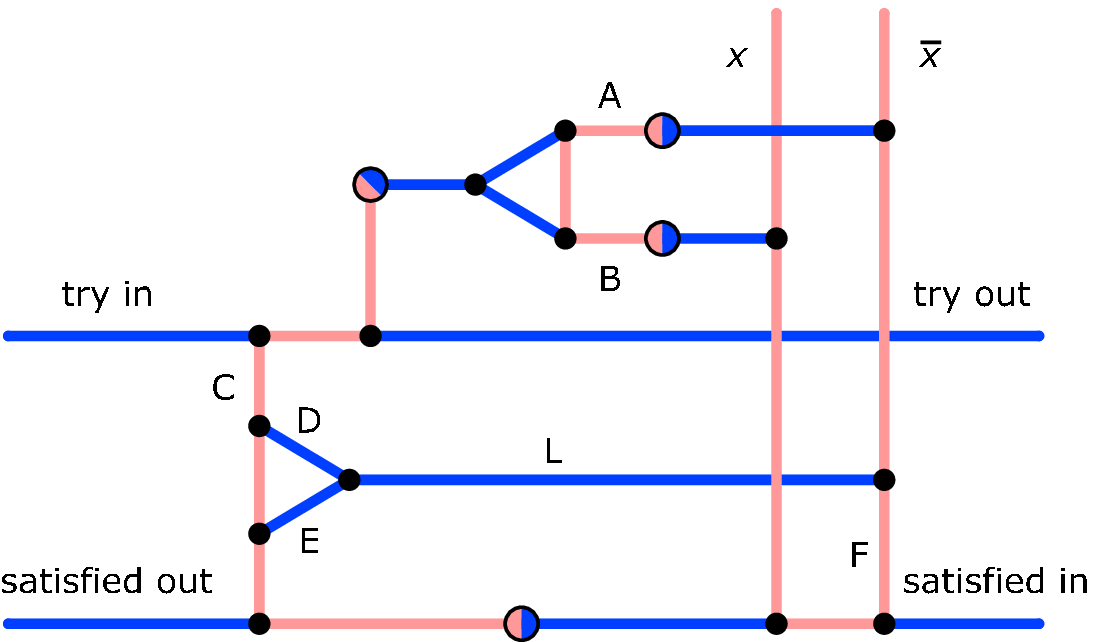}%
	\label{universal}%
}%
\caption{\label{qbf-gadgets}
  Quantifier gadgets.}
\end{figure}

\paragraph*{Existential Quantifier.}
An existential quantifier gadget (Figure~\ref{existential}) uses a latch subgraph to represent its variable,
and beyond this latch has the minimum structure needed to meet the definition of a quantifier gadget.
If the formula is true under some assignment of an existentially quantified variable,
then its quantifier gadget may lock the latch in the corresponding state, enabling \textsf{try out}
to activate, and recursively receive the \textsf{satisfied in} signal. Receiving the \textsf{satisfied in} signal simultaneously passes on the
\textsf{satisfied out} signal to the quantifier on the left.

Here we exploit the nondeterminism in the model to choose
the correct variable assignment.

\paragraph*{Universal Quantifier.}
A universal quantifier gadget is more complicated (Figure~\ref{universal}).  It may only direct
\textsf{satisfied out} leftward if the formula is true under both variable assignments. Again we use
a latch for the variable state; this time we split the variable outputs, so they can be used internally.
In addition, we use a latch internally, as a memory bit to record that
one variable assignment has been successfully tried.
With this bit set, if the other assignment is then successfully tried, \textsf{satisfied out}  is allowed
to point out.

\begin{lem}
\label{universalworks}
A universal quantifier gadget may direct its \textsf{satisfied out} edge outward if
and only if at one time its \textsf{satisfied in} edge is directed inward while its
variable state is locked in the false ($\overline{x}$) assignment, and at a
later time the \textsf{satisfied in} edge is directed inward while its variable state is locked in the true
($x$) assignment, with \textsf{try in} directed inward throughout.
\end{lem}

\begin{pf}
First we argue that, with \textsf{try in} directed outward, edge
    \textsf{E} must point right.  The \textsf{try out} edge must be
    directed inward in this case, so by Lemma \ref{trysatlemma}, \textsf{satisfied in}
    must be directed outward.  As a consequence, \textsf{F} must point down, and thus
    \textsf{L} must point right.  On the other hand, \textsf{C} must point
    up and thus \textsf{D} must point left.  Therefore, \textsf{E} is forced
    to point right in order to satisfy its \textsc{Or} vertex.

Suppose that \textsf{try in} is directed inward, the variable is locked in the false state (edge \textsf{A} points right),
and \textsf{satisfied in} is directed inward. These conditions allow the internal latch
to be unlocked, by directing edge \textsf{L} left. With the latch unlocked, edge \textsf{E} is free to point
left. The latch may then lock again, leaving \textsf{E} pointing left (because \textsf{C} may now point down, allowing
\textsf{D} to point right).
Now, the entire edge reversal sequence that occurred between directing \textsf{try out} outward and unlocking the
internal latch may be reversed. After
\textsf{try out} has deactivated, the variable may be unlocked, and change state. 
Then, suppose
that \textsf{satisfied in} activates with the variable locked in the true state
(edge \textsf{B} points right). This condition, along with edge \textsf{E} pointing left,
is both necessary and sufficient to direct \textsf{satisfied out} outward.
\end{pf}

We summarize the behavior of both types of quantifiers with the
following property:

\begin{lem}
\label{quantifierswork}
A quantifier gadget may direct its \textsf{satisfied out} edge out if and only if its 
\textsf{try in} edge is directed in, and
the formula read from
the corresponding quantifier to the right is true given the variable assignments that are
fixed by the quantifier gadgets to the left.
\end{lem}

\begin{pf}
By induction. By Lemmas \ref{trysatlemma} and \ref{universalworks}, if a quantifier gadget's
\textsf{satisfied in} edge is directed inward and the above condition is inductively assumed, then its
\textsf{satisfied out} edge may be directed outward only if the condition is true for this quantifier gadget as well.
For the rightmost quantifier gadget, the precondition is explicitly satisfied by Lemma \ref{cnflemma} and the construction in Figure~\ref{qbf}.
\end{pf}

\begin{thm}
\label{ncl-theorem}
Configuration-to-edge is PSPACE-complete, even when the constraint graph is restricted to an \textsc{And}/\textsc{Or} graph.
\end{thm}

\begin{pf}
The graph is easily seen to have a legal configuration with the quantifier \textsf{try in} edges all directed leftward.
We start with the graph in some such configuration. 
Because of the blue loop edge attached to the leftmost quantifier's \textsf{try in} edge, we may direct that edge
rightward and activate its quantifier.
By Lemma \ref{quantifierswork}, the \textsf{satisfied out} edge of the leftmost quantifier gadget may be directed leftward
if and only if $\phi$ is true. Therefore,
deciding whether that edge may reverse also decides the QBF problem,
so configuration-to-edge is PSPACE-hard.

Configuration-to-edge is in PSPACE because the state of the constraint graph can be described in
a linear number of bits, specifying the direction of each edge, and because the
list of possible moves from any state can be computed in polynomial time.  Thus
we can nondeterministically traverse the state space, at each step
nondeterministically choosing a move to make, and maintaining the current state
but not the previously visited states.  Savitch's Theorem \cite{Savitch-1970}
says that this NPSPACE algorithm can be converted into a PSPACE algorithm.
\end{pf}

\begin{cor}
\label{ncl-corollary}
Edge-to-edge and configuration-to-configuration are PSPACE-complete, even when the constraint graph is restricted to an \textsc{And}/\textsc{Or} graph.
\end{cor}

\begin{pf}
For edge-to-edge, we use the leftmost \textsf{try in} edge as the input edge; then, as
described above, there is a legal initial configuration. Again, we use the
leftmost \textsf{satisfied out} edge as the target edge.

For configuration-to-configuration, we start in some configuration with the leftmost
\textsf{try in} edge directed left, and with the four edges attached to the
output subgraph in Figure~\ref{qbf} directed as indicated. Then this same configuration, but with those four
edges reversed, is reachable if and only if $\phi$ is true. This subgraph is actually a latch: directing \textsf{satisfied out} left
allows those edges to reverse. Then \textsf{satisfied out} can be directed right again, and the entire move sequence can be
reversed.

The same algorithm as above serves to show that these tasks are also in PSPACE.
\end{pf}

\subsection{Planar Nondeterministic Constraint Logic}
\label{PNCL}

The result obtained in the previous section used particular
constraint graphs, which turn out to be nonplanar.
Thus, reductions from NCL to other problems must provide a way to encode arbitrary 
graph connections into their particular structure. 
For 2D motion-planning kinds of
problems, such a reduction would typically require some kind of crossover gadget.
Crossover gadgets are a common requirement in complexity results for these kinds
of problems, and can be among the most difficult gadgets to design.
For example, the
crossover gadget used in the proof that Sokoban is PSPACE-complete
\cite{Culberson-1998} is quite intricate.
A crossover gadget is also among those used in the Rush Hour proof
\cite{Flake-Baum-2002}.

In this section we show that any \textsc{And}/\textsc{Or} constraint graph can be translated into an
equivalent planar \textsc{And}/\textsc{Or} constraint graph (with respect to the three 
decision problems),
obviating the need for crossover gadgets in reductions from NCL.
\begin{figure}
\centering
\subfigure[Crossover]
{
	\includegraphics[width=.45\linewidth]{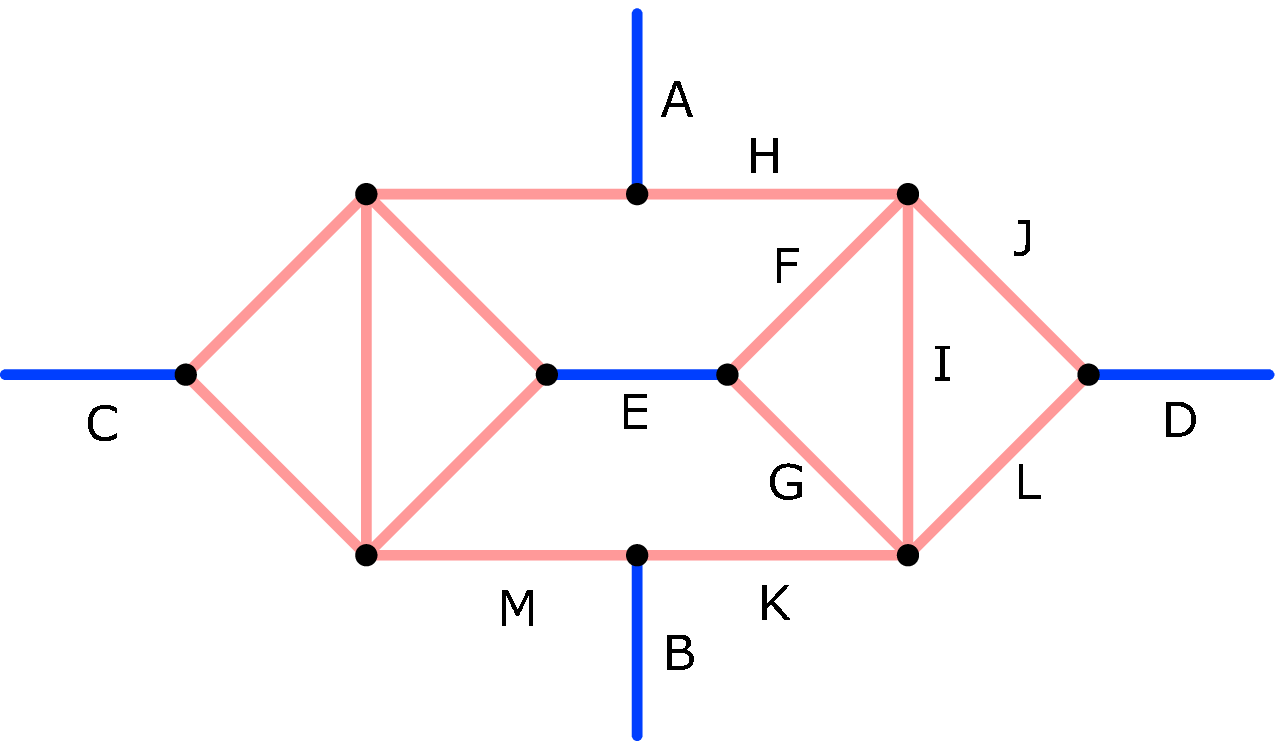}
	\label{crossover}
}\hfil
\subfigure[Half-crossover]
{
	\includegraphics[width=.32\linewidth]{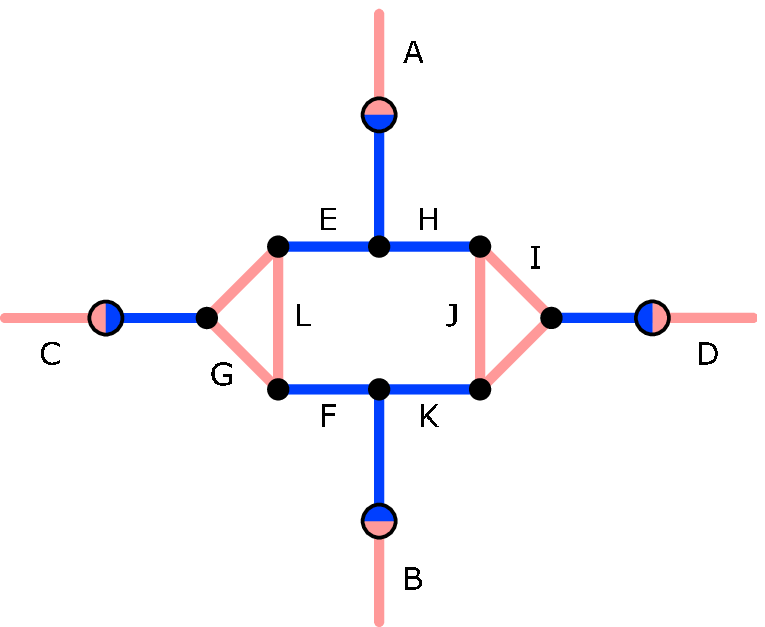}
	\label{half-crossover}
}
\caption{Planar crossover gadgets.}
\end{figure}

Figure~\ref{crossover} illustrates the reduction.  In addition to \textsc{And} and \textsc{Or}
vertices, this subgraph contains \mbox{red-red-red-red} vertices; these need any two edges to be directed
inward. (Next we will show how to substitute \textsc{And}/\textsc{Or} subgraphs for these vertices.)

\begin{lem}
\label{crossover-lemma}
In a crossover subgraph, each of the edges \textsf{A} and \textsf{B} may face outward
if and only if the other faces inward, and each of the edges \textsf{C} and \textsf{D}
may face outward if and only if the other faces inward.
\end{lem}

\begin{pf}
We show that edge \textsf{B} can
face down if and only if \textsf{A} does, and \textsf{D} can face right if and only if
\textsf{C} does.  Then by symmetry, the reverse relationships also hold.

Suppose \textsf{A} faces up, and assume without loss of generality that \textsf{E} faces
left.  Then so do \textsf{F}, \textsf{G}, and \textsf{H}.  Because \textsf{H} and \textsf{F} face left,
\textsf{I} faces up.  Because \textsf{G} and \textsf{I} face up, \textsf{K} faces
right, so \textsf{B} must face up.
Next, suppose \textsf{D} faces right, and assume without loss of generality that \textsf{I}
faces down.  Then \textsf{J} and \textsf{F} must face right, and therefore so must \textsf{E}.  An
identical argument shows that if \textsf{E} faces right, then so does \textsf{C}.

Suppose \textsf{A} faces down. Then \textsf{H} may face right, \textsf{I} may face down, and \textsf{K} may face
left (because \textsf{E} and \textsf{D} may not face away from each other).
Symmetrically, \textsf{M} may face right; therefore \textsf{B} may face down.
Next, suppose \textsf{D} faces left, and assume without loss of generality that \textsf{B}
faces up.  Then \textsf{J} and \textsf{L} may face left, and \textsf{K} may face right.
Therefore \textsf{G} and \textsf{I} may face up.  Because \textsf{I} and \textsf{J} may face up,
 \textsf{F} may face left; therefore, \textsf{E} may face left.  An identical
argument shows that \textsf{C} may also face left.
\end{pf}

Next, we must show how to represent the degree-4 vertices in
Figure~\ref{crossover} with equivalent \textsc{And}/\textsc{Or} subgraphs. The necessary subgraph
is shown in Figure~\ref{half-crossover}. Note that this subgraph implicitly contains red-blue conversions
subgraphs (Figure~\ref{red-to-blue}); we must be careful to keep the graph planar
when joining their free red edges.
We join these edges in pairs: \textsf{A}'s to \textsf{D}'s, and \textsf{B}'s to \textsf{C}'s.

\begin{lem}
\label{half-crossover-lemma}
In a half-crossover gadget, at least two of the edges \textsf{A}, \textsf{B}, \textsf{C}, 
and \textsf{D} must face inward; any two may face outward.
\end{lem}

\begin{pf}
Suppose that three edges face outward. Without loss of generality, assume that they
include \textsf{A} and~\textsf{C}. Then \textsf{E} and \textsf{F} must face left. This forces
\textsf{H} to face left and \textsf{I} and \textsf{J} to face up; then \textsf{D} must face left and
\textsf{K} must face right. But then \textsf{B}  must face up,
contradicting the assumption.

Next we must show that any two edges may face outward. We already showed how to
face \textsf{A} and \textsf{C} outward. \textsf{A} and \textsf{B} may face outward if \textsf{C} and \textsf{D}
face inward: we may face \textsf{G} and \textsf{L} down, \textsf{F} and \textsf{K} right, \textsf{I} and \textsf{J}
up, and \textsf{H} and \textsf{E} left, satisfying all vertex constraints. Also, \textsf{C} and \textsf{D} may face
outward if \textsf{A} and \textsf{B} face inward; the obvious orientations satisfy all the constraints.
By symmetry, all of the other cases are also possible.
\end{pf}

The crossover subgraph has blue free edges; what if we need to cross red edges, or a red and a blue edge?
For crossing red edges, we may attach red-blue conversion subgraphs to the crossover subgraph in two pairs, as
we did for the half-crossover. We may avoid having to cross a red edge and a blue edge, as follows: replace
one of the blue edges with a blue-red-blue edge sequence, using two red-blue conversion subgraphs, with their
free red edges joined. Then the original blue edge may be effectively crossed by crossing two red edges instead.

\begin{thm}
\label{planar}
Theorem~\ref{ncl-theorem} and Corollary~\ref{ncl-corollary} remain valid even when the input \textsc{And}/\textsc{Or}
constraint graphs are planar.

\end{thm}

\begin{pf}
Lemmas  \ref{crossover-lemma} and \ref{half-crossover-lemma}.
Any crossing edge pairs
may be replaced by the above constructions; a crossing edge may be reversed if and only if
a corresponding crossover edge (e.g., \textsf{A} or \textsf{C}) may be reversed. For two of the
decision problems, we must also
specify configurations in the replacement graph corresponding to source or target configurations,
but this is easy: pick any legal configuration of the crossover subgraphs with matching crossover edges.
\end{pf}

\subsection{Protected O{\fakesmallcaps R} Graphs}

So far we have shown that the decision problems for planar \textsc{And}/\textsc{Or} constraint graphs are PSPACE-complete.
We can make the conditions required for PSPACE-completeness still weaker; this will make our following puzzle reductions
simpler.

\begin{figure}
\centering
\includegraphics[scale=0.8]{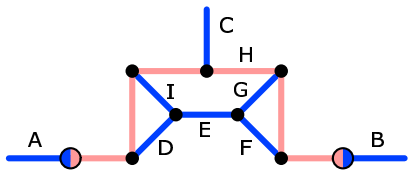}
\caption{\label{protected-or}
\textsc{Or} vertex made with protected \textsc{Or} vertices.%
}
\end{figure}

We call an \textsc{Or} vertex \emph{protected} if there are two of its edges that, due to global constraints, cannot simultaneously
be directed inward.
Intuitively, graphs with only protected \textsc{Or}s are easier to reduce to another problem domain, since the corresponding \textsc{Or}
gadgets need not function correctly in all the cases that a true \textsc{Or} must. We can simulate an \textsc{Or} vertex with a subgraph all of whose \textsc{Or} vertices are protected, as shown in Figure~\ref{protected-or}.

\begin{lem}
\label{protected-or-lemma}
Edges \textsf{A}, \textsf{B}, and \textsf{C} in Figure~\ref{protected-or} satisfy the same constraints as an O{\fakesmallcaps R} vertex;
all O{\fakesmallcaps R}s in this subgraph are protected.
\end{lem}

\begin{pf}
Suppose that edges \textsf{A} and \textsf{B} are directed outward. Then \textsf{D} and \textsf{F} must be directed
away from \textsf{E}. Assume without loss of generality that \textsf{E} points left. Then so must \textsf{G}; this forces
\textsf{H} right and \textsf{C} down, as required. 
Then, if \textsf{B} points left, the following move sequence is possible (moves on unlabeled edges omitted):
\textsf{I} right, \textsf{E} right, \textsf{G} right, \textsf{H} left, \textsf{F} left, \textsf{E} left, \textsf{I} left, \textsf{C} up. 
Similarly, we can direct \textsf{A} in, and \textsc{B} and \textsc{C} out.

The two \textsc{Or} vertices shown in the subgraph are protected: edges \textsf{I} and \textsf{D} cannot both be directed inward,
due to the red edge they both touch; similarly, \textsf{G} and \textsf{F} cannot both be directed inward. The red-blue conversion
subgraphs also contain \textsc{Or} vertices, but these are also protected. (We connect the free red edges from the conversion
subgraphs together, preserving planarity.)
\end{pf}

\begin{thm}
\label{protected-or-thm}
Theorem~\ref{planar} remains valid even when all of the \textsc{Or} vertices in the input graph are protected.
\end{thm}

\begin{pf}
Lemma  \ref{protected-or-lemma}.
Any \textsc{Or} vertex
may be replaced by the above construction; an \textsc{Or} edge may be reversed if and only if
a corresponding subgraph edge (\textsf{A}, \textsf{B}, or \textsf{C}) may be reversed. For two of the
decision problems, we must also
specify configurations in the replacement graph corresponding to source or target configurations:
pick any legal configuration of the subgraphs with matching edges.
\end{pf}

\subsection{Nondeterministic Constraint Logic on a Polyhedron}
\label{polyhedron}

In this section we give a reduction from NCL to a particularly simple geometric form.  (This result is presented for its own sake, and is not needed for our further reductions.) 
We show that any \textsc{And}/\textsc{Or} constraint graph can be translated into an equivalent
simple planar 3-connected graph.
Steinitz's Theorem \cite{Steinitz-Rademacher-1976,Ziegler-1995-Steinitz}
says that a simple planar 3-connected graph is isomorphic to
the edges of a convex polyhedron in 3D.  Therefore, any NCL problem can be
thought of as an edge redirection problem on a convex polyhedron.

\begin{figure}
\centering
\subfigure[Adding a spur to a red edge]
{\begin{minipage}{0.4\linewidth}
    \hfil
	\includegraphics[scale=.8]{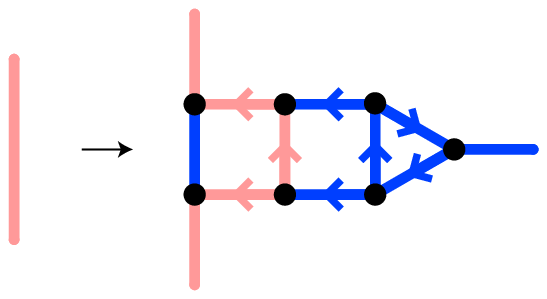}
	\label{red-to-spur}
    \hfil
 \end{minipage}
}\hfil
\subfigure[Adding a spur to a blue edge]
{\begin{minipage}{0.4\linewidth}
    \hfil
	\includegraphics[scale=.8]{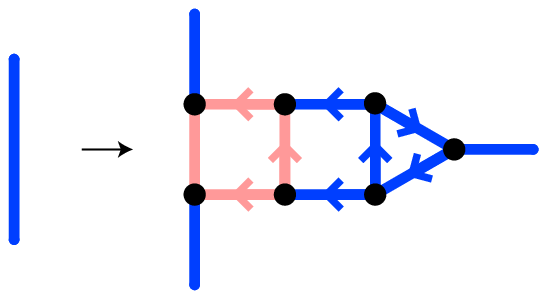}
	\label{blue-to-spur}
    \hfil
 \end{minipage}
}
\caption{\label{3-connectivity}
  3-connectivity method.}
\end{figure}

We use the constructions in Figure~\ref{3-connectivity} to perform the conversion.
We may replace any red edge with a subgraph yielding an extra unconstrained blue
edge, as shown in Figure~\ref{red-to-spur}: the original red edge may be reversed in the original
graph if and only if the top (equivalently bottom) red edge may be reversed in the new graph.
(This is like performing two consecutive red-blue conversions, but with an extra blue edge.)
Likewise, we may replace any blue edge with a similar subgraph, as shown in
Figure~\ref{blue-to-spur}.

\begin{thm}
Every \textsc{And}/\textsc{Or} constraint graph has an equivalent simple planar 3-connected
\textsc{And}/\textsc{Or} graph which can be computed in polynomial time.
\end{thm}

\begin{pf}
First, we make the graph planar, by Theorem \ref{planar}.

Next, we make the graph simple: if any two vertices are joined by more than one edge, we may
replace one with a subgraph from Figure~\ref{3-connectivity}. 
We also choose an arbitrary edge on a common face
  with the replaced edge, and replace this edge as well with such a subgraph,
  and join the two free blue edges.
We eliminate loop edges similarly. 

Suppose the resulting graph is not 3-connected. Then there exist zero, one, or two vertices which, if removed,
would separate the graph into multiple pieces. 
From each of two such pieces, choose an edge that lies on a common
face, replace these edges
with subgraphs from Figure~\ref{3-connectivity}, and join the two
free blue edges. 
(This step preserves planarity and simplicity.)
Now these pieces will not be separated by the vertex removal.
By repeating this process, we may make the graph simple, planar, and 3-connected.

As in Theorem \ref{planar}, we must also specify a mapping from original to modified graph configurations; again, we simply map orientations of the replaced edges to consistent subgraph configurations.
\end{pf}

\section{Applications}
\label{Applications}

In this section, we apply our results from the previous section to various
puzzles and motion-planning problems.
One result (sliding blocks) is completely new, and provides a tight bound;
one result (Rush Hour) reproduces an existing result, with a simpler construction; the last
result (Sokoban) strengthens an existing result.

\subsection{Sliding Blocks}
\label{sliding-block-proof}

We define the \emph{Sliding Blocks} problem as follows:
given a configuration of rectangles (\emph{blocks}) of constant sizes in a rectangular $2$-dimensional
box, can the blocks be translated and rotated, without intersection among the objects, so as to
move a particular block?

We are interested in the difficulty of this problem, for various allowed integral block sizes.
We give a reduction from configuration-to-edge for protected \textsc{Or} graphs showing that Sliding Blocks is PSPACE-hard
even when all the blocks
are $1 \times 2$ rectangles (dominoes).
(Somewhat simpler constructions are possible if larger blocks are allowed.)
In contrast, there is a simple polynomial-time algorithm for
$1 \times 1$ blocks; thus, our results are tight.

The \emph{Warehouseman's Problem} \cite{Hopcroft-Schwartz-Sharir-1984} is a related
problem in which there are no restrictions on block size, and the goal is to achieve a
particular total configuration. Its PSPACE-hardness also follows from our result.

\begin{figure}
\centering
{
	\includegraphics[width=.4\linewidth]{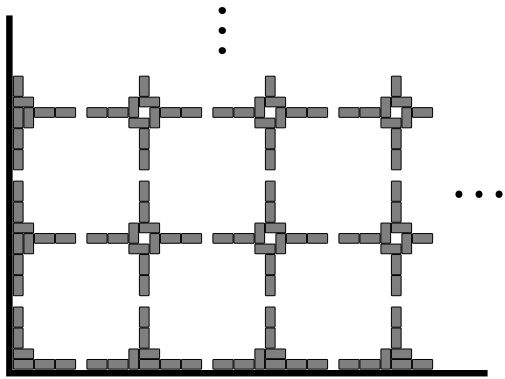}
}
\caption{\label{slide-layout}
  Sliding Blocks layout.}
\end{figure}

\paragraph*{Sliding Blocks Layout.}

We fill the box with a regular grid of gate gadgets, within a ``cell wall'' construction as
shown in Figure~\ref{slide-layout}. The internal construction of the gates is such that none
of the cell-wall blocks may move, thus providing overall integrity to the configuration.

\paragraph*{A{\fakesmallcaps ND} and O{\fakesmallcaps R} Vertices.}

\begin{figure*}
\centering
\subfigure[\textsc{And}]
{
	\includegraphics[width=.35\linewidth]{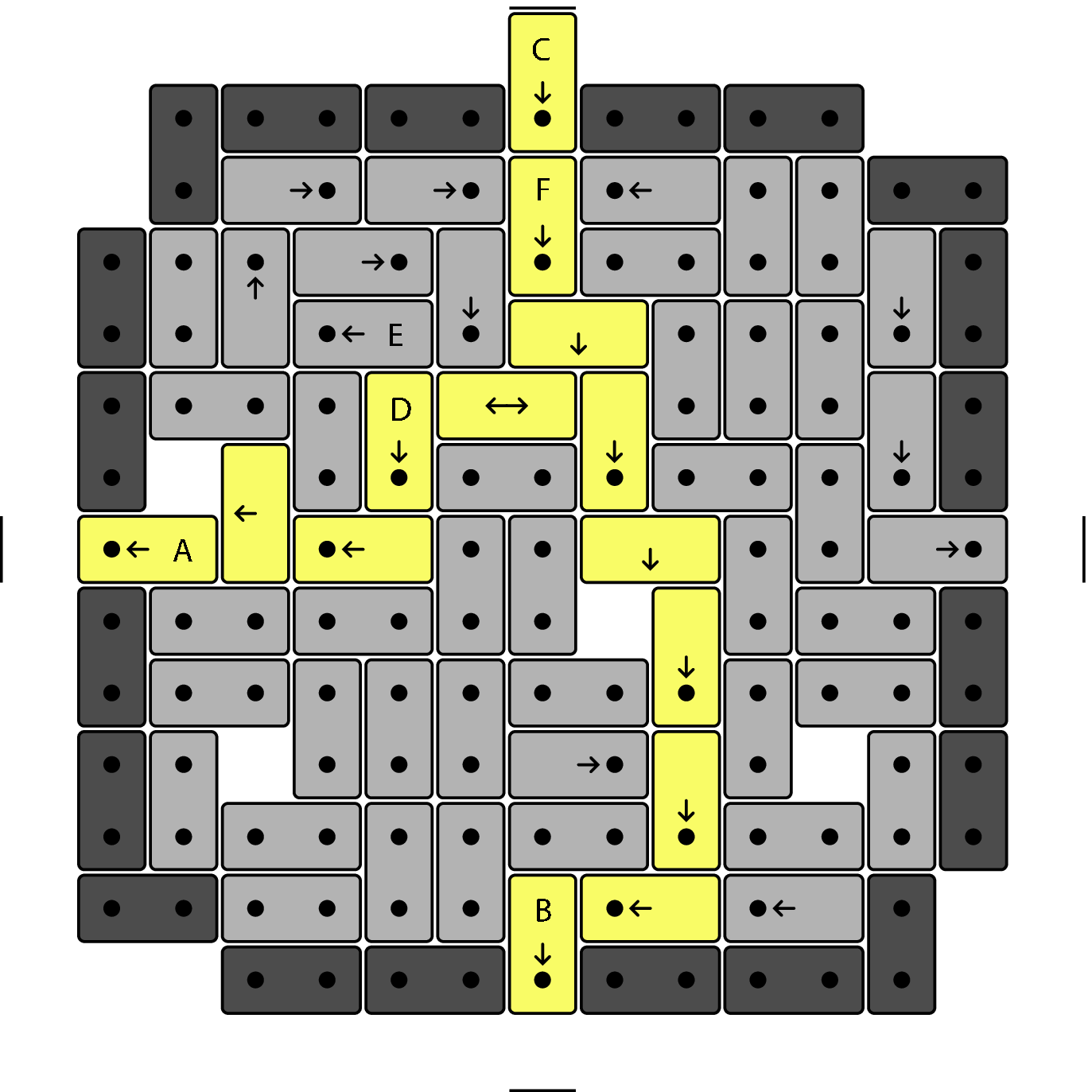}
	\label{block-and}
}\hfil\hfil
\subfigure[Protected \textsc{Or}]
{
	\includegraphics[width= .35\linewidth]{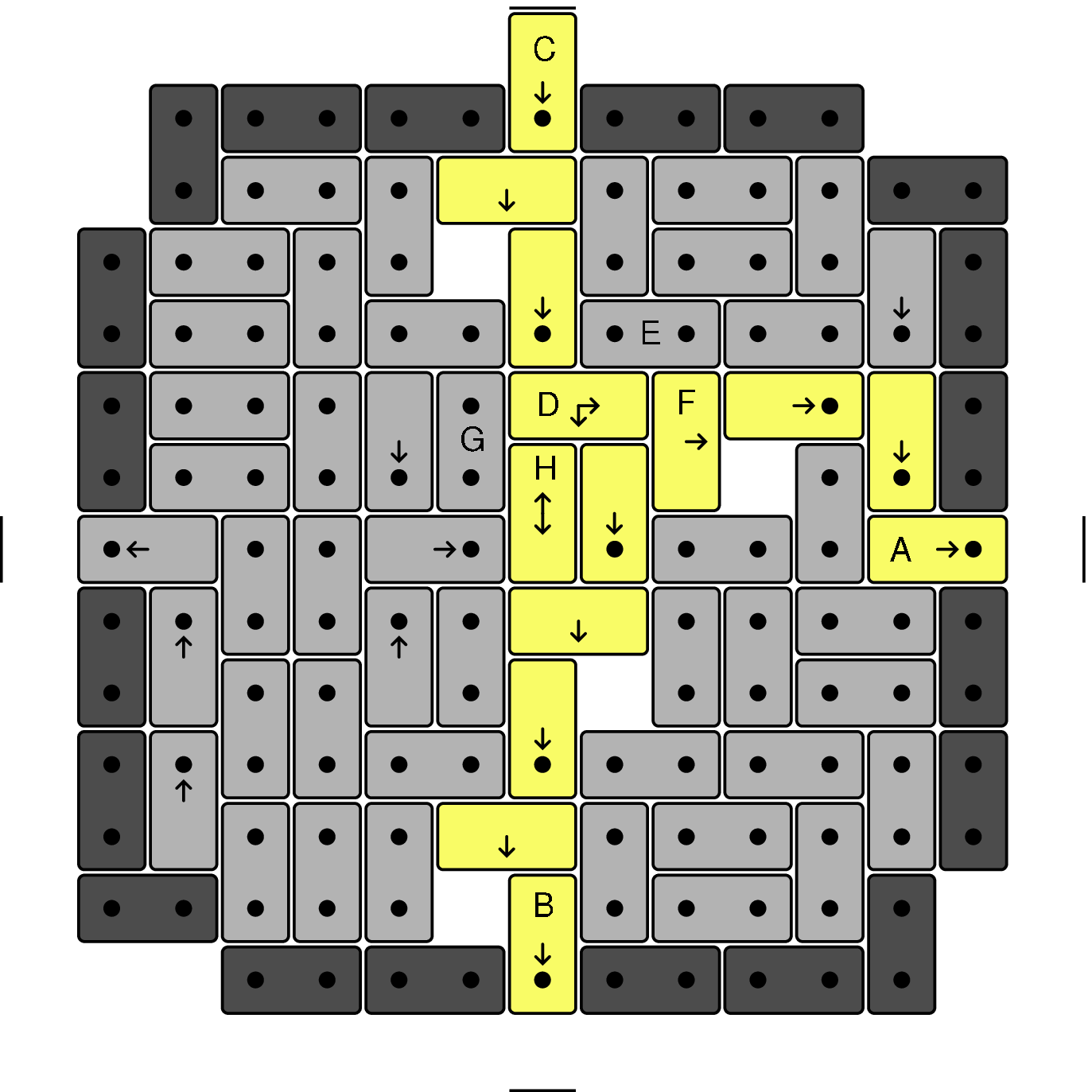}
	\label{block-or}
}
\caption{Sliding Blocks vertex gadgets.}
\end{figure*}

We construct NCL \textsc{And} and protected \textsc{Or} vertex gadgets out of dominoes,
in Figures~\ref{block-and} and \ref{block-or}. Each figure provides the bulk of an inductive proof
of its own correctness, in the form of annotations. A dot indicates a square that is always occupied;
the arrows indicate the possible positions a block can be in. For example, in
Figure~\ref{block-or}, block \textsf{D} may occupy its initial position, the position one unit to
the right, or the position one unit down (but not, as we will see, the position one unit down and one unit
right). Because we allow continuous motions, all intermediate block positions are also
possible, but this is irrelevant to the vertex properties. We also note that the constructions are
such that no block is ever free to rotate.

For each vertex gadget, we show that the annotations are correct, by inductively assuming for each
block that its surrounding annotations are correct; its correctness will then follow.
The few exceptions are noted below.
The annotations were generated by a computer search
of all reachable configurations, but are easy to verify by inspection.

In each diagram, we assume that the cell-wall blocks (dark colored) may not move outward; we
then need to show they may not move inward. The light-colored (``trigger'') blocks are the ones
whose motion serves to satisfy the vertex constraints; the medium-colored blocks are fillers.
Some of them may move, but none may move in such a way as to disrupt the vertices'
correct operation.

The short lines outside the vertex ports indicate constraints due to adjoining vertices; none of the
``port'' blocks may move entirely out of its vertex. For it to do so, the adjoining vertex would have to
permit a port block to move entirely inside the vertex, but in each diagram the annotations show
this is not possible. Note that the port blocks are shared between adjoining vertices, as are the
cell-wall blocks.
For example,
if we were to place a protected \textsc{Or} above an \textsc{And}, its bottom port block would be the
same as the \textsc{And}'s top port block.

A protruding port block corresponds to an inward-directed edge; a retracted block corresponds to
an outward-directed edge. Signals propagate by moving ``holes'' forward.
Sliding a block \emph{out} of a vertex gadget
thus corresponds to directing an edge \emph{in} to a graph vertex.

\begin{lem}
The construction in Figure~\ref{block-and} satisfies the same constraints as an NCL \textsc{And}
vertex, with \textsf{A} and \textsf{B} corresponding to the  \textsc{And} red edges, and \textsf{C} to the blue edge.
\end{lem}

\begin{pf}
We need to show that block \textsf{C} may move down if and only if block \textsf{A} first moves
left and block \textsf{B} first moves down.

First, observe that this motion is possible. The trigger blocks may each shift one unit in an
appropriate direction, so as to free block \textsf{C}.

The annotations in this case serve as a complete proof of their own correctness, with one
exception. Block~\textsf{D} appears as though it might be able to slide upward, because block
\textsf{E} may slide left, yet \textsf{D} has no upward arrow. However, for \textsf{E} to slide left,
\textsf{F} must first slide down, but this requires that \textsf{D} be first be slid down.
So when \textsf{E} slides left, \textsf{D} is not in a position to fill the space it
vacates.

Given the annotations' correctness, it is easy to see that it is not possible for \textsf{C} to
move down unless \textsf{A} moves left and \textsf{B} moves down.
\end{pf}

\begin{lem}
The construction in Figure~\ref{block-or} satisfies the same constraints as an NCL
protected \textsc{Or} vertex, with \textsf{A} and \textsf{B} corresponding to the protected edges.
\end{lem}

\begin{pf}
We need to show that block \textsf{C} may move down if and only if block \textsc{A} first
moves right, or block \textsf{B} first moves down.

First, observe that these motions are possible. If \textsf{A} moves right, \textsf{D} may move
right, releasing the blocks above it. If \textsf{B} moves down, the entire central column may
also move down.

The annotations again provide the bulk of the proof of their own correctness. In this case there
are three exceptions. Block \textsf{E} looks as if it might be able to move down, because \textsf{D}
may move down and \textsf{F} may move right. However, \textsf{D} may only move down if
\textsf{B} moves down, and \textsf{F} may only move right if \textsf{A} moves right. Because this
is a protected \textsc{Or}, we are guaranteed that this cannot happen:
the vertex will be used only
in graphs such that at most one of \textsf{A} and \textsf{B} can slide out at a time. Likewise, \textsf{G} could move right
if \textsf{D} were moved right while \textsf{H} were moved down, but again those possibilities
are mutually exclusive. Finally, \textsf{D} could move both down and right one unit, but again this
would require \textsf{A} and \textsf{B} to both slide out.

Given the annotations' correctness, it is easy to see that it is not possible for \textsf{C} to
move down unless \textsf{A} moves right or \textsf{B} moves down.
\end{pf}

\begin{figure}
\centering
\subfigure[$2 \times 2$ filler]
{
	\includegraphics[width=.11\textwidth]{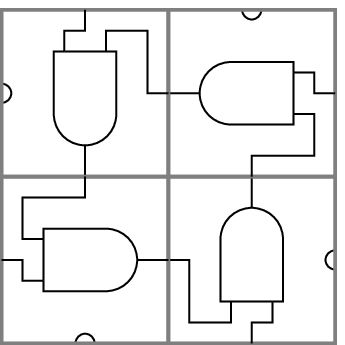}
	\label{empty-2-2}
}\hfil
\subfigure[$2 \times 3$ filler]
{
	\includegraphics[width=.11\textwidth]{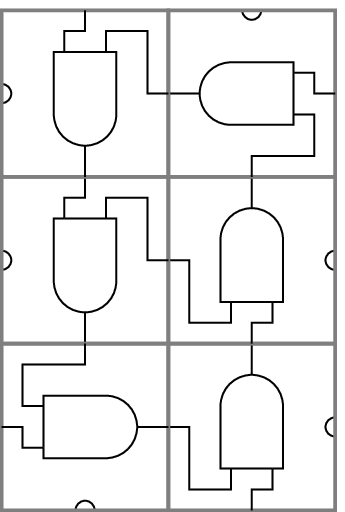}
	\label{empty-2-3}
}\hfil
\subfigure[$5 \times 5$ straight]
{
	\includegraphics[width= .19\textwidth]{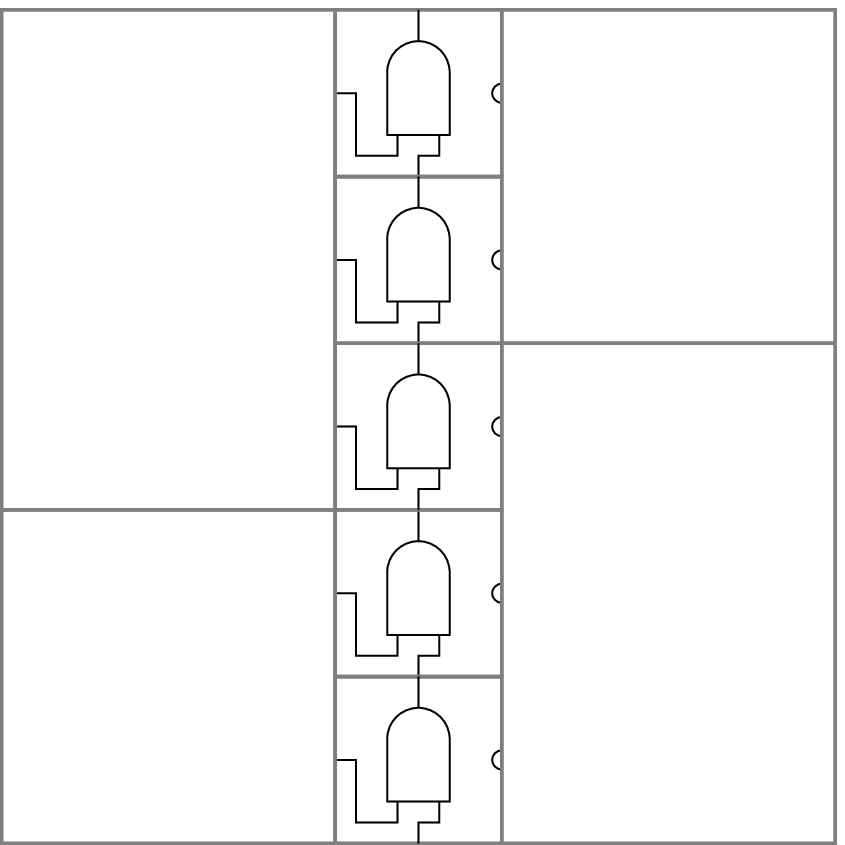}
	\label{straight-5-5}
}\hfil
\subfigure[$5 \times 5$ turn]
{
	\includegraphics[width= .19\textwidth]{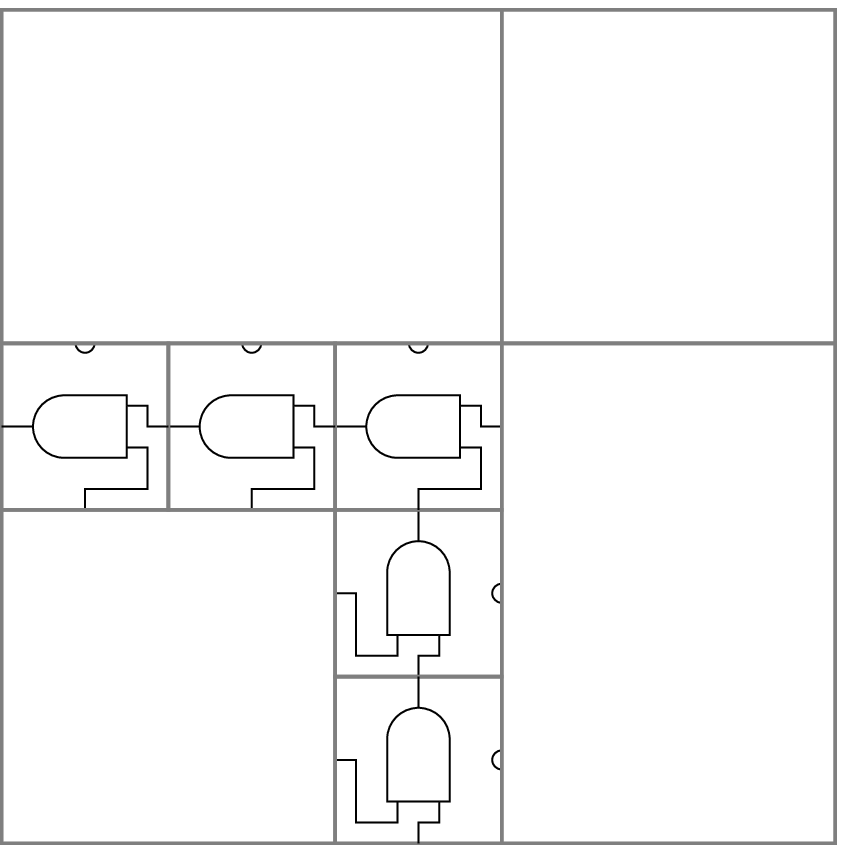}
	\label{turn-5-5}
}\hfil
\subfigure[$5 \times 5$  protected \textsc{Or}]
{
	\includegraphics[width= .19\textwidth]{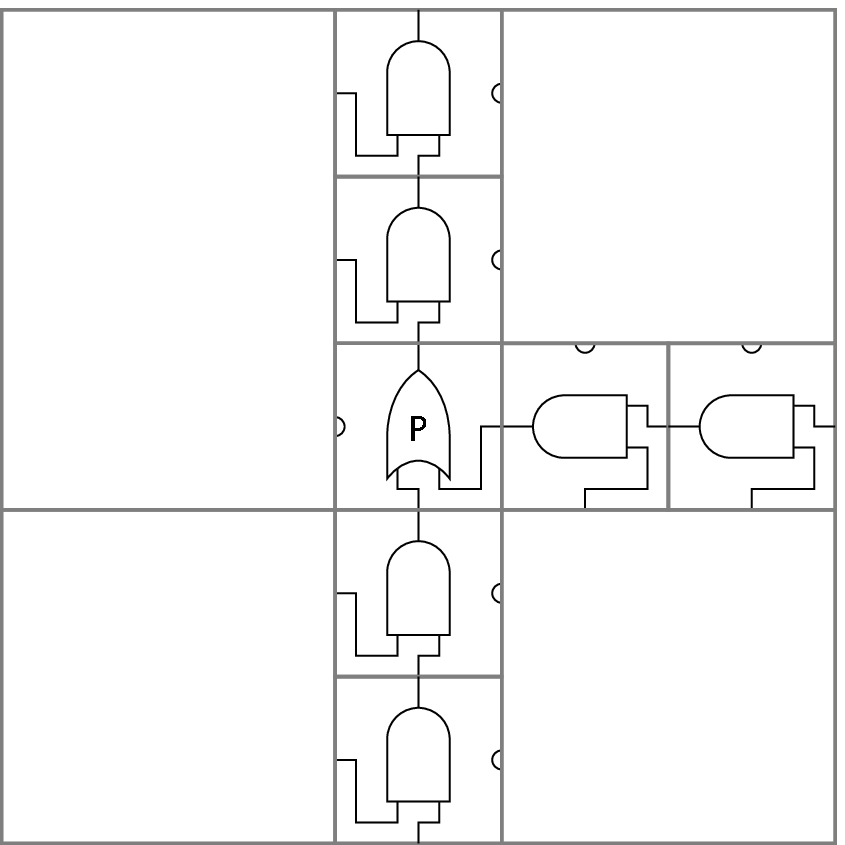}
	\label{or-5-5}
}
\caption{\label{block-wiring}
  Sliding Blocks wiring.}
\end{figure}

\paragraph*{Graphs.}
Having shown how to make \textsc{And} and protected \textsc{Or} gates out of sliding-blocks
configurations, we must now show how to connect them together into arbitrary planar
graphs.
First, note that the box wall constrains the facing port blocks of the vertices adjacent to it
to be retracted (see Figure~\ref{slide-layout}). This does not present a problem, however, as we will show.
The unused ports of both the \textsc{And}
and protected \textsc{Or} vertices are unconstrained; they may be slid in or out with no effect on the
vertices. Figures~\ref{empty-2-2} and \ref{empty-2-3} show how to make $(2 \times 2)$-vertex
and $(2 \times 3)$-vertex ``filler'' blocks out of \textsc{And}s. (We use conventional ``and'' and ``or'' icons
to denote the vertex gadgets.)
Because none of the \textsc{And}s
need ever activate, all the exterior ports of these blocks are unconstrained. (The unused
ports are drawn as semicircles.)

We may use these filler blocks to build $(5 \times 5)$-vertex blocks corresponding
to ``straight'' and ``turn'' wiring elements (Figures~\ref{straight-5-5} and \ref{turn-5-5}).
Because the filler blocks may supply the missing inputs to the \textsc{And}s, the
``output'' of one of these blocks may activate (slide in) if and only if the ``input'' is active (slid out).
Also, we may ``wrap'' the \textsc{And} and protected \textsc{Or} vertices in $5 \times 5$ ``shells'',
as shown for protected \textsc{Or} in Figure \ref{or-5-5}. (Note that ``left turn'' is the same as
``right turn''; switching the roles of input and output results in the same constraints.)

We use these $5 \times 5$ blocks to fill the layout; we may line the edges of the layout
with unconstrained ports. The straight and turn blocks provide the necessary flexibility
to construct any planar graph, by letting us extend the vertex edges around the layout as
needed.

\begin{thm}
Sliding Blocks is PSPACE-hard, even for $1 \times 2$ blocks.
\end{thm}

\begin{pf}
Reduction from configuration-to-edge for planar protected \textsc{Or} graphs, by the construction
described. A port block of a 
particular vertex gadget may move if and only if the corresponding NCL graph edge may
be reversed.
\end{pf}

\begin{cor}
The Warehouseman's Problem is PSPACE-hard, even for $1 \times 2$ blocks.
\end{cor}

\begin{pf}
As above, but using configuration-to-configuration instead of configuration-to-edge.
The NCL graph initial and desired configurations correspond to two block
configurations; the second is reachable from the first if and only if the NCL problem has a
solution.
\end{pf}

If we restrict the block motions to unit translations, then these problems are also in
PSPACE, as in Theorem \ref{ncl-theorem}.

\subsection{Rush Hour}

In the puzzle \emph{Rush Hour}, one is given a sliding-block configuration with the additional
restriction that each block is constrained to move only horizontally or vertically on a grid. The goal is
to move a particular block to a particular location at the edge of the grid.
In the commercial version of the puzzle, the grid is $6 \times 6$, the blocks are all $1 \times 2$
or $1 \times 3$ (``cars'' and ``trucks''), and each block constraint direction is the same as its lengthwise orientation.

Flake and Baum \cite{Flake-Baum-2002} showed that the generalized problem is
PSPACE-complete, by showing how to build a kind of reversible computer from
Rush Hour gadgets that work like our \textsc{And} and \textsc{Or} vertices, as well as a
crossover gadget. 
Tromp \cite{Tromp-2000-rush-hour} strengthened their result by showing that
Rush Hour is PSPACE-complete even if the blocks are all $1 \times 2$.

Here we give a simpler construction showing that Rush Hour is PSPACE-complete, again
using the traditional $1 \times 2$ and $1 \times 3$ blocks which must slide lengthwise.
We only need an \textsc{And} and a protected \textsc{Or},
which turns out to be easier to build than
\textsc{Or}; because of
our generic crossover construction (Section \ref{PNCL}), we don't need a crossover
gadget. (We also don't need the miscellaneous wiring gadgets used in \cite{Flake-Baum-2002}.)

\begin{figure}
\centering
\subfigure[Layout]
{
	\includegraphics[width=.4\textwidth]{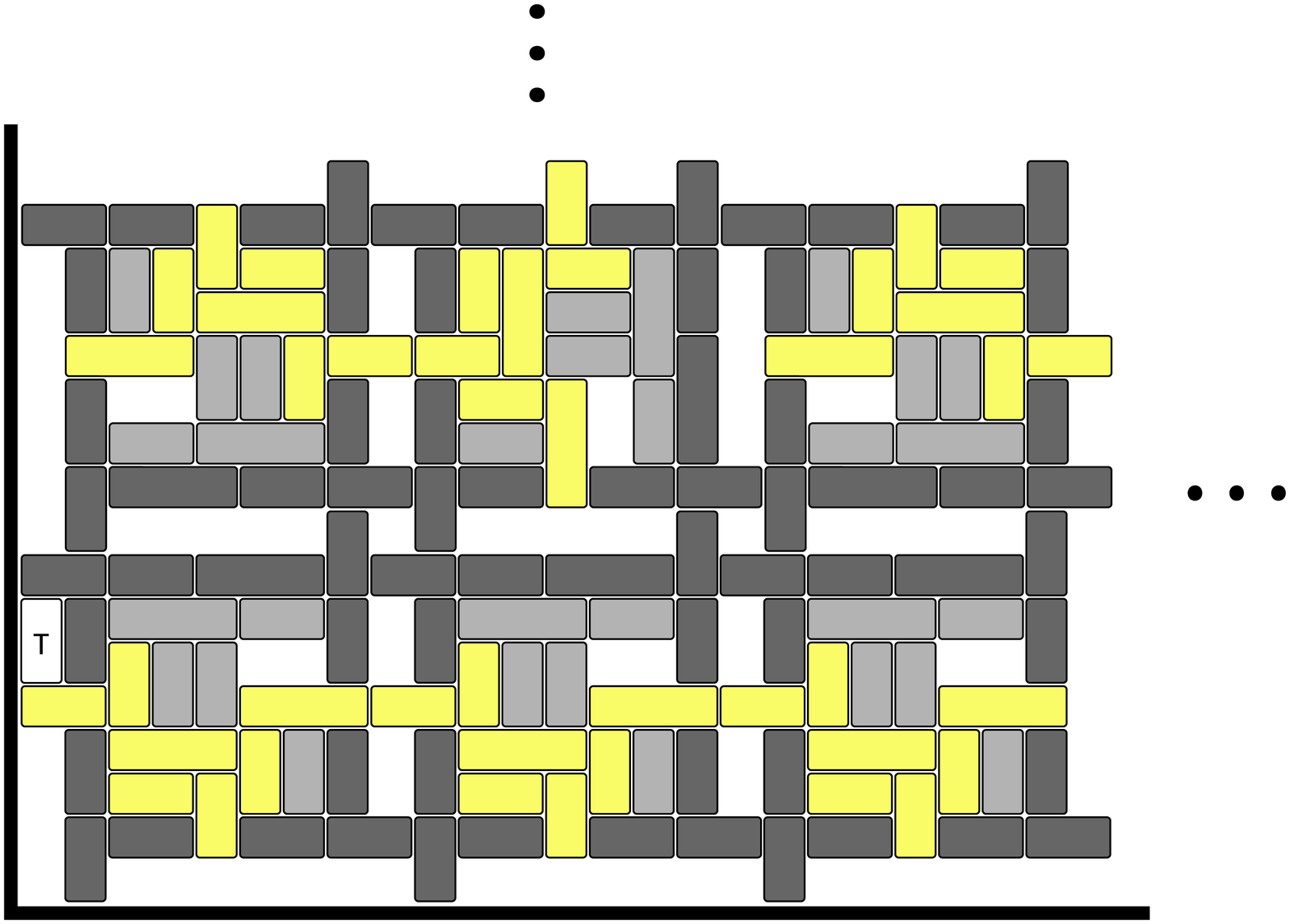}
	\label{rh-layout}
}
\hspace{.22 in}
\subfigure[\textsc{And}]
{
	\includegraphics[width=.18\textwidth]{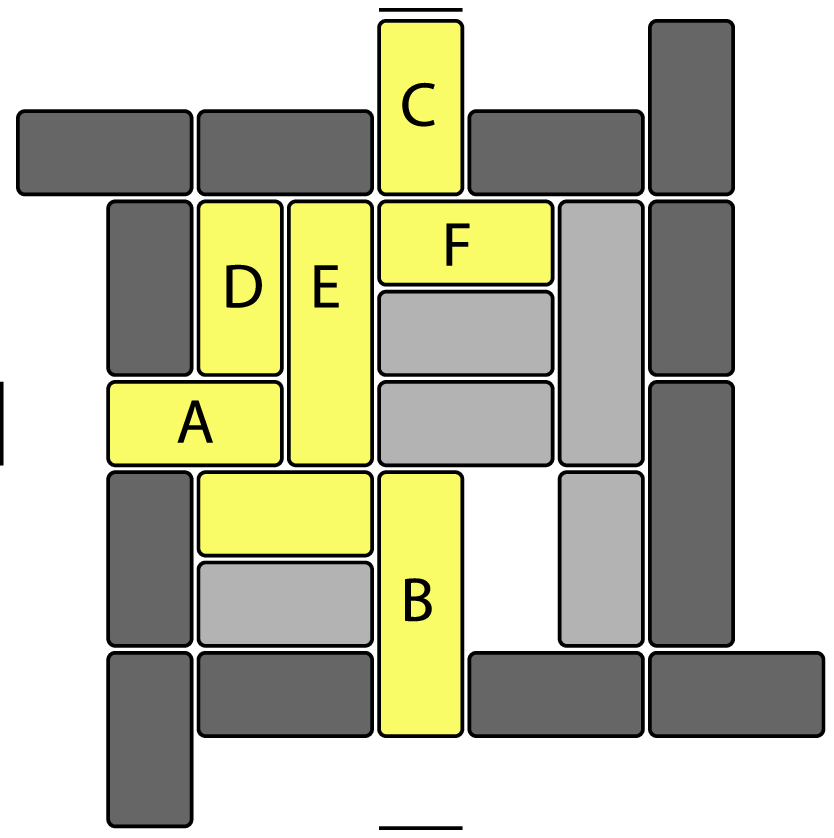}
	\label{rh-and}
}
\hspace{.22 in}
\subfigure[Protected \textsc{Or}]
{
	\includegraphics[width=.18\textwidth]{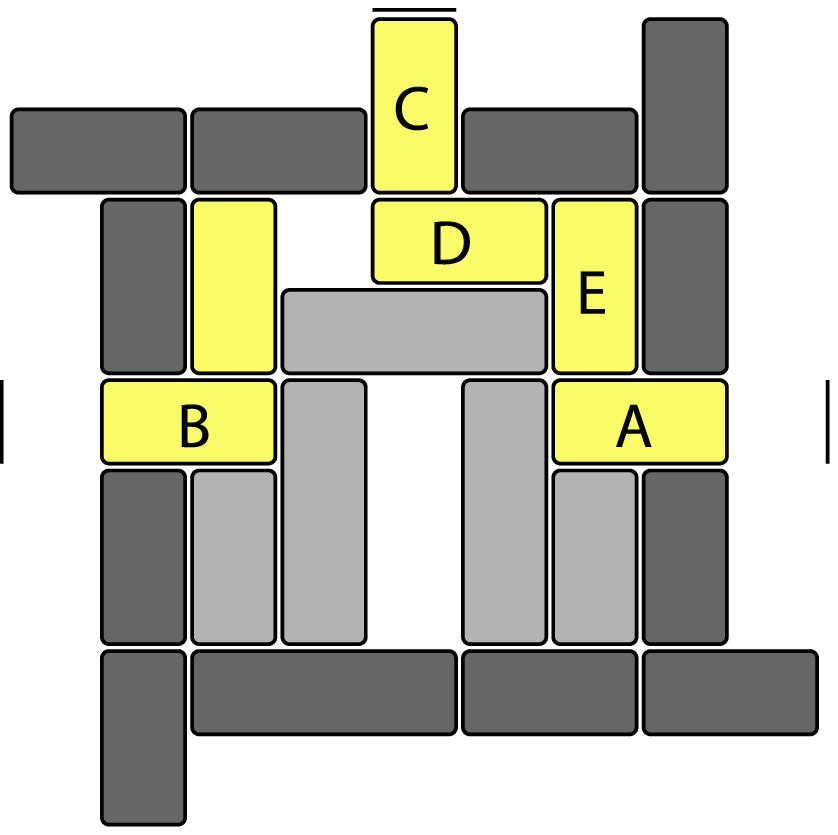}
	\label{rh-latch}
}
\caption{Rush Hour layout and vertex gadgets.}
\label{rh}
\end{figure}

\paragraph*{Rush Hour Layout.}
We tile the grid with our vertex gadgets, as shown in Figure~\ref{rh-layout}. 
One block
(\textsf{T}) is the target, which must be moved to the bottom left corner; it is released when
a particular port block slides into a vertex.

Dark-colored blocks represent the ``cell walls'', which unlike in our sliding-blocks construction
are not shared. They are arranged so that they may not move at all.
Light-colored blocks are ``trigger'' blocks, whose motion serves to satisfy
the vertex constraints. Medium-colored blocks are fillers; some of them may move, but they
don't disrupt the vertices' operation.

As in the sliding-blocks construction, edges are directed inward by sliding blocks out of the
vertex gadgets; edges are directed outward by sliding blocks in.
The layout ensures
that no port block may ever slide out into an adjacent vertex; this helps keep the
cell walls fixed.

\begin{lem}
The construction in Figure~\ref{rh-and} satisfies the same constraints as an NCL
\textsc{And} vertex, with \textsf{A} and \textsf{B} corresponding to the  \textsc{And} red edges, and \textsf{C} to the blue edge.
\end{lem}

\begin{pf}
We need to show that \textsf{C}
may move down if and only if \textsf{A} first moves left and \textsf{B} first moves down.

Moving \textsf{A} left and \textsf{B} down allows \textsf{D} and \textsf{E} to slide down, freeing \textsf{F},
which releases \textsf{C}. The filler blocks on the right ensure that \textsf{F} may only move
left; thus, the inputs are required to move to release the output.
\end{pf}

\begin{lem}
The construction in Figure~\ref{rh-latch} satisfies the same constraints as an NCL
protected \textsc{Or} vertex, with \textsf{A} and \textsf{B} corresponding to the protected edges.
\end{lem}

\begin{pf}
We need to show that \textsf{C}
may move down if either \textsf{A} first moves left or \textsf{B} first moves right.

If either \textsf{A} or \textsf{B} slides out, this allows \textsf{D} to slide out of the way of \textsf{C}, as required. Note that we are using the protected \textsc{Or} property: if \textsf{A} were to move right, \textsf{E} down, \textsf{D} right, \textsf{C} down, and \textsf{B} left,
we could not then slide \textsf{A} left, even though the \textsc{Or} property should allow this; \textsf{E} would keep \textsf{A} blocked. But in a protected \textsc{Or}, we are guaranteed that \textsf{A} and \textsf{B} will not simultaneously be slid out.
\end{pf}

\paragraph*{Graphs.}
We may use the same constructions here we used for sliding-blocks layouts: $5 \times 5$ blocks of Rush Hour vertex gadgets serve to build all
the wiring necessary to construct arbitrary planar graphs (Figure~\ref{block-wiring}).

In the special case of arranging for the target block to reach its destination, this
will not quite suffice; however, we may direct the relevant signal to the bottom
left of the grid, and then remove the bottom two rows of vertices from the bottommost
$5 \times 5$ blocks; these can have no effect on the graph. The resulting
configuration, shown in Figure~\ref{rh-layout}, allows the target block to be
released properly.

\begin{thm}
Rush Hour is PSPACE-complete.
\end{thm}

\begin{pf}
Reduction from configuration-to-edge for planar protected \textsc{Or} graphs, by the construction
described.
The output port block of a 
particular vertex may move if and only if the corresponding NCL graph edge may
be reversed. We direct this signal to the lower left of the grid, where it may release
the target block.

Rush Hour is in PSPACE: a simple nondeterministic algorithm traverses the
state space, as in Theorem \ref{ncl-theorem}.
\end{pf}

\paragraph*{Generalized Problem Bounds.}
We may consider the more general \emph{Constrained Sliding Block} problem,
where blocks need not be $1 \times 2$ or $1 \times 3$, and may have a constraint
direction independent of their dimension. In this context, the existing Rush Hour
results do not yet provide a tight bound; the complexity of the problem for
$1 \times 1$ blocks has not been addressed. 

Deciding whether a
block may move at all is in P: e.g, we may do a breadth-first search for a movable block that would ultimately enable the target block to move, beginning with the blocks obstructing the target block. Since no block need ever move more than once to free a dependent block, it is safe to terminate the search at already-visited blocks.

Therefore, a straightforward application of our technique cannot show this problem hard; however, the complexity of moving a given block to a given position
is not obvious.
Tromp and Cilibrasi \cite{Tromp-Cilibrasi-2004-rush-hour} provide some empirical indications that minimum-length solutions for $1 \times 1$ Rush Hour may grow exponentially with puzzle size. 

\subsection{Sokoban}
In the pushing-blocks puzzle \emph{Sokoban}, one is given a configuration of $1 \times 1$
blocks, and a set of target positions. One of the blocks is distinguished as the \emph{pusher}.
A move consists of moving the pusher a single unit either vertically or horizontally; if a block
occupies the pusher's destination, then that block is pushed into the adjoining space, providing
it is empty. Otherwise, the move is prohibited. Some blocks are \emph{barriers}, which may not
be pushed. The goal is to make a sequence of moves such that there is a (non-pusher) block
in each target position.

Culberson \cite{Culberson-1998} proved that Sokoban is PSPACE-complete, by showing
how to construct a Sokoban position corresponding to a space-bounded Turing machine. 
Using NCL, we give an alternate proof.  Our result applies even if there are no barriers
allowed in the Sokoban position, thus strengthening Culberson's result.

\begin{figure}
\centering
\subfigure[\textsc{And}]
{
	\includegraphics[width=.25\linewidth]{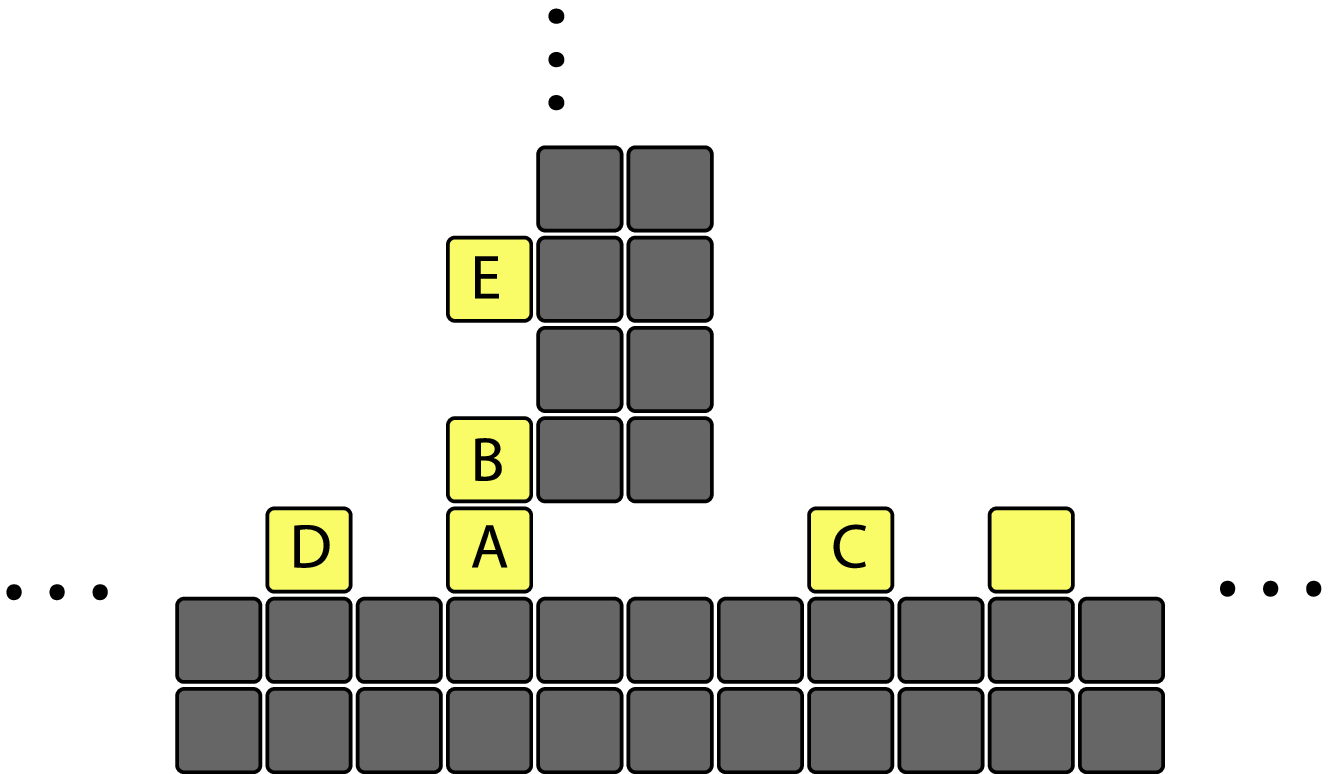}
	\label{sok-and}
}\hfil
\subfigure[\textsc{Or}]
{
	\includegraphics[width=.261\linewidth]{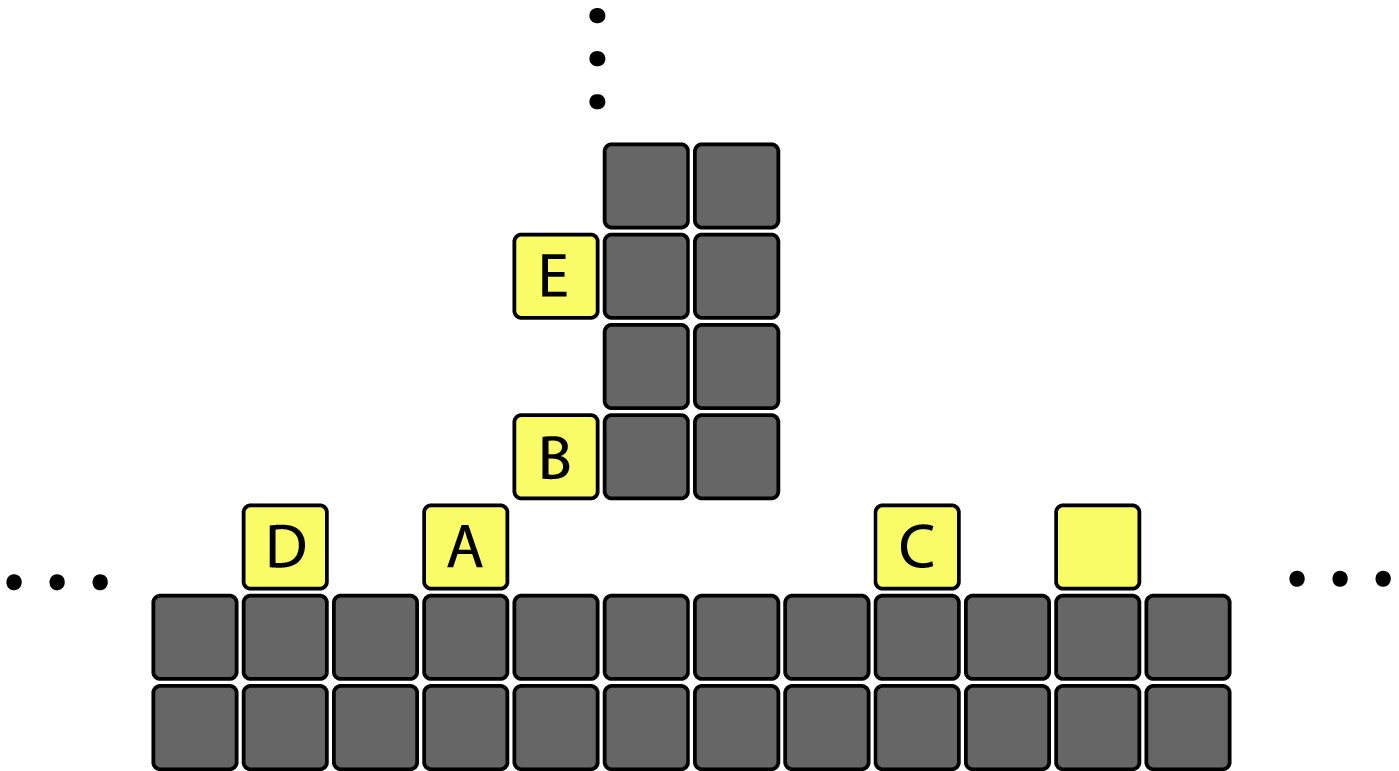}
	\label{sok-or}
}\hfil
\subfigure[Utility gadgets]
{
	\includegraphics[width=.445\linewidth]{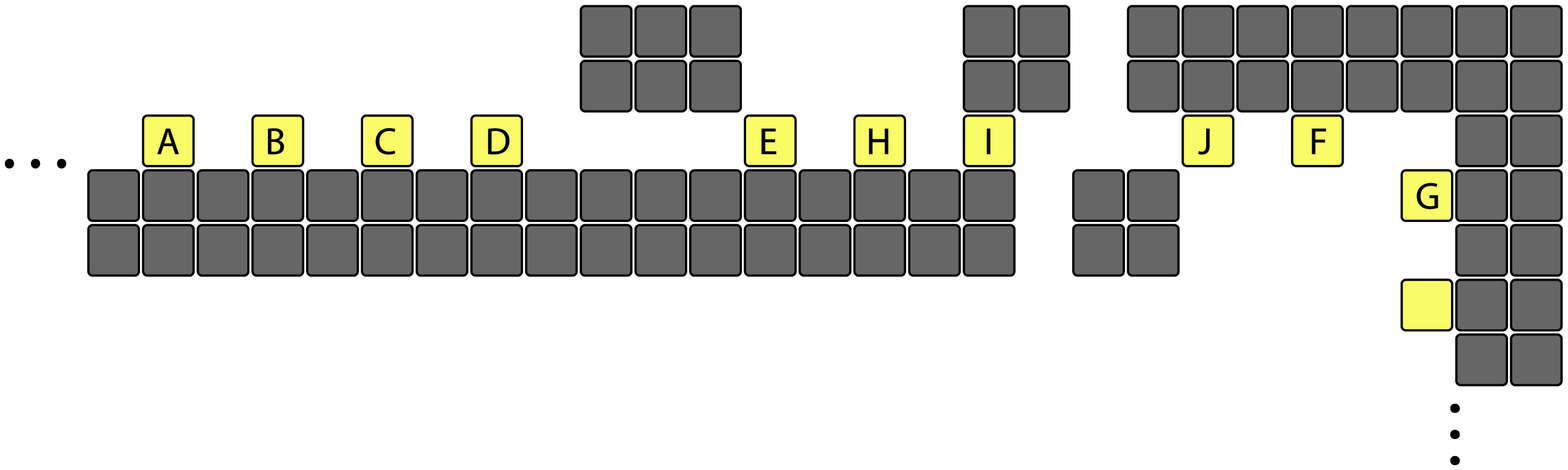}
	\label{sok-utility}
}
\caption{\label{sok-gadgets}
  Sokoban gadgets.}
\end{figure}

\paragraph*{Unrecoverable Configurations.}
The idea of an \emph{unrecoverable configuration} is central to Culberson's proof, and
it will be central to our proof as well.
We construct our Sokoban instance so that if the puzzle is solvable, then the original
configuration may be restored from any solved state by reversing all the pushes. Then any
push which may not be reversed leads to an unrecoverable configuration.
For example, in the partial configuration in Figure~\ref{sok-and},
if block \textsf{A} is pushed left, it will be irretrievably
stuck next to block~\textsf{D}; there is no way to position the pusher so as to move it again.
We may speak of such a move as being prohibited,
or impossible, in the sense that no solution to the puzzle can include such a move, even
though it is technically legal.

\paragraph*{A{\fakesmallcaps ND} and O{\fakesmallcaps R} Vertices.}
We construct NCL \textsc{And} and \textsc{Or} vertex gadgets out of partial Sokoban positions,
in Figure \ref{sok-gadgets}. (The pusher is not shown.)
The dark-colored blocks in the figures, though unmovable, are not barriers; they are simply blocks that cannot be moved by the pusher
   because of their configuration. The light-colored (``trigger'') blocks are the ones whose motion serves to satisfy the vertex
constraints.
In each vertex, blocks \textsf{A} and \textsf{B} represent outward-directed edges; block
\textsf{C} represents an inward-directed edge. \textsf{A} and \textsf{C} switch state by
moving left one unit; \textsf{B} switches state by moving up one unit.
We assume that the pusher may freely move to any empty space surrounding
a vertex. We also assume that block \textsf{D} in
Figure~\ref{sok-and} may not reversibly move left more than one unit.
Later, we show how to arrange both of these conditions.

\begin{lem}
\label{sok-and-works}
The construction in Figure~\ref{sok-and} satisfies the same constraints as an NCL \textsc{And}
vertex, with \textsf{A} and \textsf{B} corresponding to the  \textsc{And} red edges, and \textsf{C} to the blue edge.
\end{lem}

\begin{pf}
We need to show that \textsf{C} may move left if and only if \textsf{A} first
moves left, and \textsf{B} first moves up.
For this to happen, \textsf{D} must first move left, and \textsf{E} must
first move up; otherwise pushing \textsf{A} or \textsf{B} would lead to an 
unrecoverable configuration. Having first pushed \textsf{D} and \textsf{E} out of the
way, we may then push \textsf{A} left, \textsf{B} up, and \textsf{C} left.
However, if we push \textsf{C} left without first pushing \textsf{A} left and \textsf{B}
up, then we will be left in an unrecoverable configuration; there will be no way
to get the pusher into the empty space left of \textsf{C} to push it right again. (Here we
use the fact that \textsf{D} can only move left one unit.)
\end{pf}

\begin{lem}
The construction in Figure~\ref{sok-or} satisfies the same constraints as an NCL \textsc{Or}
vertex.
\end{lem}

\begin{pf}
We need to show that \textsf{C} may move left if and only if \textsf{A} first
moves left, or \textsf{B} first moves up.

As before, \textsf{D} or \textsf{E} must first move out of the way to allow \textsf{A} or \textsf{B}
to move. Then, if \textsf{A} moves left, \textsf{C} may be pushed left; the gap opened
up by moving \textsf{A} lets the pusher get back in to restore \textsf{C} later.
Similarly for~\textsf{B}.

However, if we push \textsf{C} left without first pushing \textsf{A} left or \textsf{B}
up, then, as in Lemma \ref{sok-and-works}, we will be left in an unrecoverable
configuration.
\end{pf}

\paragraph*{Graphs.}
We have shown how to make \textsc{And} and \textsc{Or} vertices, but we must
still show how to connect them up into arbitrary planar graphs. The remaining
gadgets we shall need are illustrated in Figure~\ref{sok-utility}.

The basic idea is to connect the vertices together with alternating sequences of
blocks placed against a double-thick wall, as in the left of Figure~\ref{sok-utility}.
Observe that for block \textsf{A} to move right, first \textsf{D} must move right, then \textsf{C},
then \textsf{B}, then finally \textsf{A}, otherwise two blocks will wind up stuck together.
Then, to move block \textsf{D} left again, the reverse sequence must occur.
Such movement sequences serve to propagate activation from one vertex
to the next.

We may switch the ``parity'' of such strings, by interposing an appropriate group of six
blocks: \textsf{E} must move right for \textsf{D} to, then \textsf{D} must move back left for \textsf{E} to.
We may turn corners: for \textsf{F} to move right, \textsf{G} must first move down.
Finally, we may ``flip'' a string over, to match a required orientation at the next
vertex, or to allow a turn in a desired direction: for \textsf{H} to move right, \textsf{I} must 
move right at least two spaces; this requires that \textsf{J} first move right.

We satisfy the requirement that block \textsf{D} in Figure~\ref{sok-and} may not reversibly move left
more than one unit by protecting the corresponding edge of every \textsc{And} with a turn; observe that 
in Figure~\ref{sok-utility}, block \textsf{F} may not reversibly move right more than one unit.
The flip gadget solves our one remaining problem: how to position the
pusher freely wherever it is needed. Observe that it is always possible for the
pusher to cross a string through a flip gadget. (After moving \textsf{J} right, we may
actually move \textsf{I} \emph{three} spaces right.) If we simply place at least one
flip along each wire, then the pusher can get to any side of any vertex.

\begin{thm}
Sokoban is PSPACE-complete, even if no barriers are allowed.
\end{thm}

\begin{pf}
Reduction from configuration-to-configuration for planar \textsc{And}/\textsc{Or} graphs.
Given a planar \textsc{And}/\textsc{Or} graph, we build a Sokoban
puzzle as described above, corresponding to the initial graph configuration. 
We place a target at every position that would be occupied by
a block in the Sokoban configuration corresponding to the target graph configuration.
Since NCL is inherently reversible, and our construction
emulates NCL, then the solution configuration must also be reversible,
as required for the unrecoverable configuration constraints.

Sokoban is in PSPACE: a simple nondeterministic algorithm traverses the
state space, as in Theorem \ref{ncl-theorem}.
\end{pf}

\section{Alternative Formulation}
\label{Tokens}

In this section we give an alternative formulation of the NCL problem, \emph{Sliding Tokens}, in terms of tokens sliding along graph edges. This formulation is even simpler than the edge-reversal formulation in Section \ref{our model}. However, it lacks the inherent computational flavor of A{\fakesmallcaps ND}/O{\fakesmallcaps R} constraint graphs; furthermore, it seems to be less suitable for reductions. Therefore, we have chosen to use A{\fakesmallcaps ND}/O{\fakesmallcaps R} constraint graphs as our primary formulation; this section may be viewed as an additional application.

Our ``machine'' in this case is a graph. A configuration of a machine is a subset of its vertices containing tokens, such that no two tokens are adjacent along an edge. A move from one configuration to another is made by moving a token from one vertex to an adjacent one, resulting in a valid configuration. The decision question is whether a given token can eventually be moved by a sequence of moves.

We give a reduction from configuration-to-edge for A{\fakesmallcaps ND}/O{\fakesmallcaps R} graphs showing that this problem is PSPACE-complete.

\paragraph*{A{\fakesmallcaps ND}/O{\fakesmallcaps R} Graphs.}

\begin{figure*}
\centering
\subfigure[\textsc{And}]
{
	\includegraphics[scale=.5]{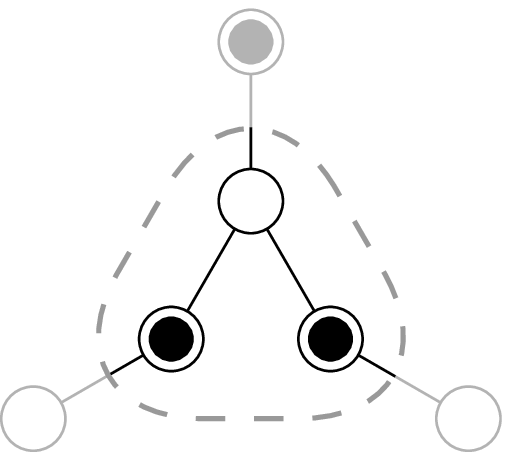}
	\label{token-and}
}\hfil\hfil
\subfigure[\textsc{Or}]
{
	\includegraphics[scale=.5]{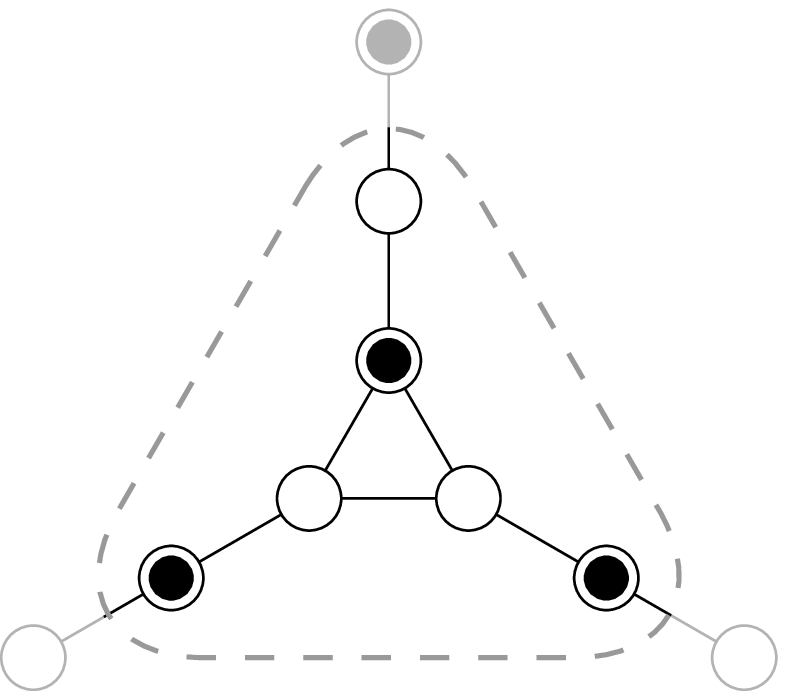}
	\label{token-or}
}
\caption{Sliding Tokens vertex gadgets.}
\label{token-gadgets}
\end{figure*}

We construct NCL A{\fakesmallcaps ND} and O{\fakesmallcaps R} vertex gadgets out of sliding-token subgraphs, in Figures \ref{token-and} and \ref{token-or}. The edges that cross the dotted-line gadget borders are ``port'' edges.
A token on an outer port-edge vertex represents an inward-directed NCL edge, and vice-versa.
Given an A{\fakesmallcaps ND}/O{\fakesmallcaps R} graph and configuration, we construct a corresponding sliding-token graph, by joining together A{\fakesmallcaps ND} and O{\fakesmallcaps R} vertex gadgets at their shared port edges, placing the port tokens appropriately.

\begin{thm}
Sliding Tokens is PSPACE-complete.
\end{thm}

\begin{pf}
First, observe that no port token may ever leave its port edge. Choosing a particular port edge \textsf{E}, if we inductively assume that this condition holds for all other port edges, then there is never a legal move outside \textsf{E} for its token -- another port token would have to leave its own edge first.

The \textsc{And} gadget clearly satisfies the same constraints as an NCL \textsc{And} vertex; the upper token can slide in just when both lower tokens are slid out.
Likewise, the upper token in the \textsc{Or} gadget can slide in when either lower token is slid out -- the internal token can then slide to one side or the other to make room. It thus satisfies the same constraints as an NCL \textsc{Or} vertex.

Sliding Tokens is in PSPACE: a simple nondeterministic algorithm traverses the
state space, as in Theorem \ref{ncl-theorem}.
\end{pf}

\paragraph*{Discussion.} This problem formulation is interesting for several reasons. It is a dynamic version of the Independent Set problem, which is NP-complete \cite{Garey-Johnson-1979}. Similarly, the natural two-player-game version of Independent Set, called Kayles, is also PSPACE-complete \cite{Garey-Johnson-1979}. Just as many NP-complete problems become PSPACE-complete when turned into two-player games \cite{Schaefer-1976}, it is also natural to expect that they become PSPACE-complete when turned into dynamic puzzles.

From a recreational standpoint, sliding-token graphs are similar both to sliding-coin puzzles and to $1 \times 1$ sliding-block puzzles, many forms of which are in P \cite{AlgGameTheoryMFCS2001}. Typically one needs some structure of the pieces beyond an atomic token or $1 \times 1$ block to add complexity to a motion-planning puzzle. In this case, however, the nonadjacency requirement suffices.

Computationally, sliding-token graphs also superficially resemble Petri nets.

\section{Conclusion}
\label{Conclusion}

We proved that one of the simplest possible forms of motion planning,
sliding $1 \times 2$ blocks (dominoes) around in a box,
is PSPACE-hard. 
This result is a major strengthening of previous
results.
The problem has no artificial constraints, such as the
movement restrictions of Rush Hour; it has object size constraints which
are tightly bounded, unlike the unbounded object sizes in the Warehouseman's Problem.
Also compared to the Warehouseman's Problem, the task is simply to move a block
at all, rather than to reach a total configuration.

Along the way, we presented a model of computation of interest in its own
right, and which can be used to prove several motion-planning problems to be
PSPACE-hard.  Our hope is to apply this approach to several other
motion-planning problems whose complexity remain open, for example:

\begin{enumerate}

  \item \textbf{$1 \times 1$ Rush Hour.}  While $1 \times 1$ sliding blocks
    can be solved in polynomial time, if we enforce horizontal or vertical
    motion constraints as in Rush Hour, does the problem become
    PSPACE-complete \cite{Tromp-Cilibrasi-2004-rush-hour}?
  Deciding whether a block may move at all
    is in P, so a straightforward application of our technique will not work,
    but what is the complexity of moving a given block to a given position?
  \item \textbf{Retrograde Chess.}  Given two configurations of chess pieces
    in a generalized $n \times n$ board, is it possible to play from one
    configuration to the other if the players cooperate?
    This problem is known to be NP-hard \cite{Bodlaender-2001-retrograde};
    is it PSPACE-complete?
\end{enumerate}

\section*{Acknowledgments}

We thank John Tromp, Shafira Goldwasser, and Albert Meyer for several useful comments and suggestions.
We also thank Timothy Chow for introducing us to the Retrograde Chess problem
mentioned in the conclusion, and for pointing out the reference
\cite{Bodlaender-2001-retrograde}.

\bibliography{combinatorialgames,complexity,motionplanning,polytopes,pushingblocks}
\bibliographystyle{plain}

\end{document}